\documentclass[useAMS,usenatbib,usegraphicx]{mn2e}
\usepackage{graphicx}
\usepackage[breaklinks,colorlinks,citecolor=black,linkcolor=black,urlcolor=black]{hyperref}

\newcommand*{\mysub}[2]{\ensuremath{#1_{\mathrm{#2}}}}
\newcommand*{\unit}[1]{\ensuremath{\mathrm{\, #1}}}

\newcommand*{\Msun}{\ensuremath{\, M_{\odot}}}
\newcommand*{\mproton}{\mysub{m}{p}}
\newcommand*{\keV}{\unit{keV}}
\newcommand*{\erg}{\unit{erg}}
\newcommand*{\cm}{\unit{cm}}
\newcommand*{\km}{\unit{km}}
\newcommand*{\Mpc}{\unit{Mpc}}
\newcommand*{\second}{\unit{s}}

\newcommand*{\NH}{\mysub{N}{H}}
\newcommand*{\Mgas}{\mysub{M}{gas}}
\newcommand*{\Mlens}{\mysub{M}{lens}}
\newcommand*{\mgas}{\mysub{m}{gas}}
\newcommand*{\mlens}{\mysub{m}{lens}}
\newcommand*{\fgas}{\mysub{f}{gas}}

\newcommand*{\Yx}{\mysub{Y}{X}}
\newcommand*{\dA}{\mysub{d}{A}}

\newcommand*{\LCDM}{\ensuremath{\Lambda}CDM}
\newcommand*{\Omegam}{\mysub{\Omega}{m}}
\newcommand*{\Omegab}{\mysub{\Omega}{b}}

\newcommand*{\rhocr}{\mysub{\rho}{cr}}

\newcommand*{\E}[1]{\ensuremath{\times 10^{#1}}}
\newcommand*{\expectation}[1]{\ensuremath{\left\langle #1 \right\rangle}}
\newcommand*{\ltsim}{\ {\raise-.75ex\hbox{$\buildrel<\over\sim$}}\ }
\newcommand*{\gtsim}{\ {\raise-.75ex\hbox{$\buildrel>\over\sim$}}\ }
\newcommand*{\proptosim}{\ {\raise-.75ex\hbox{$\buildrel\propto\over\sim$}}\ }

\newcommand*{\Chandra}{{\it Chandra}}
\newcommand*{\Planck}{{\it Planck}}

\defcitealias{Mantz0909.3098}{M10a}
\newcommand*{\mare}{\citetalias{Mantz0909.3098}}
\defcitealias{Mantz0909.3099}{M10b}
\newcommand*{\maerd}{\citetalias{Mantz0909.3099}}
\defcitealias{von-der-Linden1208.0597}{I}
\newcommand*{\pone}{\citetalias{von-der-Linden1208.0597}}
\defcitealias{Kelly1208.0602}{II}

\defcitealias{Applegate1208.0605}{III}
\newcommand*{\pthree}{\citetalias{Applegate1208.0605}}
\defcitealias{Mantz1407.4516}{IV}
\newcommand*{\pfour}{\citetalias{Mantz1407.4516}}
\defcitealias{Mantz1502.06020}{M15}
\newcommand*{\morph}{\citetalias{Mantz1502.06020}}

\newcommand*{\figscale}{0.9}

\begin{document}

\title[Weighing the Giants V: Cluster Scaling Relations]{Weighing the Giants V: Galaxy Cluster Scaling Relations}

\author[A. B. Mantz et al.]{Adam B. Mantz,$^{1,2}$\thanks{E-mail: \href{mailto:amantz@slac.stanford.edu}{\tt amantz@slac.stanford.edu}} {}
  Steven W. Allen,$^{1,2,3}$
  R. Glenn Morris,$^{1,3}$\newauthor
  Anja von der Linden,$^{1,2,4}$
  Douglas E. Applegate,$^5$
  Patrick L. Kelly,$^6$\newauthor
  David L. Burke,$^{1,3}$
  David Donovan,$^{7}$
  Harald Ebeling$^{7}$
  \smallskip
  \\$^1$Kavli Institute for Particle Astrophysics and Cosmology, Stanford University, 452 Lomita Mall, Stanford, CA 94305, USA
  \\$^2$Department of Physics, Stanford University, 382 Via Pueblo Mall, Stanford, CA 94305, USA
  \\$^3$SLAC National Accelerator Laboratory, 2575 Sand Hill Road, Menlo Park, CA  94025, USA
  \\$^4$Department of Physics and Astronomy, Stony Brook University, Stony Brook, NY 11794, USA
  \\$^5$Argelander-Institute for Astronomy, Auf dem H\"ugel 71, D-53121 Bonn, Germany
  \\$^6$Department of Astronomy, University of California, Berkeley, CA 94720, USA
  \\$^{7}$Institute for Astronomy, 2680 Woodlawn Drive, Honolulu, HI 96822, USA
}
\date{Submitted 10 June 2016}

\pagerange{\pageref{firstpage}--\pageref{lastpage}} \pubyear{????}
\maketitle
\label{firstpage}

\begin{abstract}
  We present constraints on the scaling relations of galaxy cluster X-ray luminosity, temperature and gas mass (and derived quantities) with mass and redshift, employing masses from robust weak gravitational lensing measurements. These are the first such results obtained from an analysis that simultaneously accounts for selection effects and the underlying mass function, and directly incorporates lensing data to constrain total masses. Our constraints on the scaling relations and their intrinsic scatters are in good agreement with previous studies, and reinforce a picture in which departures from self-similar scaling laws are primarily limited to cluster cores. However, the data are beginning to reveal new features that have implications for cluster astrophysics and provide new tests for hydrodynamical simulations. We find a positive correlation in the intrinsic scatters of luminosity and temperature at fixed mass, which is related to the dynamical state of the clusters. While the evolution of the nominal scaling relations over the redshift range $0.0<z<0.5$ is consistent with self similarity, we find tentative evidence that the luminosity and temperature scatters respectively decrease and increase with redshift. Physically, this likely related to the development of cool cores and the rate of major mergers. We also examine the scaling relations of redMaPPer richness and Compton $Y$ from \Planck{}. While the richness--mass relation is in excellent agreement with recent work, the measured $Y$--mass relation departs strongly from that assumed in the \Planck{} cluster cosmology analysis. The latter result is consistent with earlier comparisons of lensing and \Planck{} scaling-relation-derived masses.
\end{abstract}


\section{Introduction} \label{sec:intro}

The scaling relations of galaxy clusters, stochastic functions describing the dependence of observables on mass and redshift, provide a phenomenological and statistical description of the astrophysical diversity in the evolving cluster population (see the review of \citealt{Giodini1305.3286}). Processes such as cooling, galaxy and star formation, feedback, and hierarchical growth through merging all play some role in determining the scatter and overall trends of quantities measured at X-ray, optical and mm wavelengths. Additionally, the scaling relations play a central role in obtaining cosmological constraints from cluster number counts, linking theoretical predictions of the mass function to the observables used to define complete cluster samples from surveys \citep*{Allen1103.4829}.

Perhaps the most important recent addition to the studies of cluster scaling relations and cosmology is the ability to robustly estimate cluster total masses using measurements of the weak gravitational lensing of background galaxies. Although the intrinsic scatter between true mass and lensing mass is larger than some other mass proxies (e.g.\ gas mass and temperature), lensing has the advantage of being much simpler to understand theoretically than proxies based on gas physics and, when systematics are sufficiently well controlled, is expected to be unbiased at the per cent level \citep{Becker1011.1681}. As a result of the intrinsic scatter and the limited size of current weak lensing follow-up samples, the primary (and not inconsiderable) impact of lensing data so far has been to provide an unbiased calibration for the normalizations of scaling relations, and thereby of the matter power spectrum (see Paper~\pfour{} in this series, \citealt{Mantz1407.4516}). Nevertheless, as weak lensing measurements for massive clusters are the closest tool we currently have to a ``true'' mass measurement, testing whether the slopes and scatters of scaling relations derived using other proxies are consistent with lensing data represents a major milestone for the field.

In order to infer the scaling relations of the cluster population as introduced above, as opposed to the purely empirical trends in an observed sample, features of the data such as the selection function and the underlying mass function must be accounted for. In detail, the required analysis is identical to the one used to constrain cosmological parameters from cluster number counts and follow-up data; the astrophysical and cosmological models of interest are degenerate, and, in general, should be fitted simultaneously (\citealt{Mantz0909.3098}, hereafter \mare{}; \citealt*{Allen1103.4829}). This was our approach in Paper~\pfour{}, which primarily concerned the cosmological constraints enabled by the robust lensing mass estimates presented in Papers~\pone{}--\pthree{} of this series \citep{von-der-Linden1208.0597, Kelly1208.0602, Applegate1208.0605}. Here we present results based on the same analysis method, but focussing on the scaling relations and their astrophysical implications. To this end, we incorporate a larger amount of X-ray follow-up measurements, reduced using more up-to-date calibration information, than was included in Paper~\pfour{}. The additional X-ray data, and the use of external cosmological data to provide constraints on cosmological parameters, allow us to explore and discuss various generalizations of the ``baseline'' scaling model used in that work.

In Section~\ref{sec:data}, we describe the analysis of \Chandra{} and ROSAT X-ray data employed here, and the resulting measurements of gas masses, temperatures and X-ray luminosities. Section~\ref{sec:model} reviews the models for the scaling relations, and for the follow-up observations used to constrain them. Our results on X-ray scaling relations are presented in Section~\ref{sec:results} and their astrophysical consequences are discussed in Section~\ref{sec:discussion}. Section~\ref{sec:szo} presents scaling relations of optical richness and Compton $Y$, which we fit using relatively simple methods (not accounting for selection effects and cosmological degeneracies). We conclude in Section~\ref{sec:conclusion}. We follow the convention of defining characteristic cluster masses and radii in terms of a spherical overdensity, $\Delta$, with respect to the critical density of the Universe at a given cluster's redshift, $M_\Delta = (4/3) \pi \Delta \rhocr(z) r_\Delta^3$. Unless otherwise specified, quoted best fits and uncertainties refer to marginalized posterior modes and 68.3 per cent confidence maximum-probability intervals, and plotted quantities are derived assuming a flat \LCDM{} cosmological model with Hubble parameter $h=H_0/100\km\second^{-1}\Mpc^{-1}=0.7$ and mean matter density $\Omegam=0.3$.

\vspace{-5mm}
\section{Data} \label{sec:data}

\begin{figure}
  \centering
  \includegraphics[scale=\figscale]{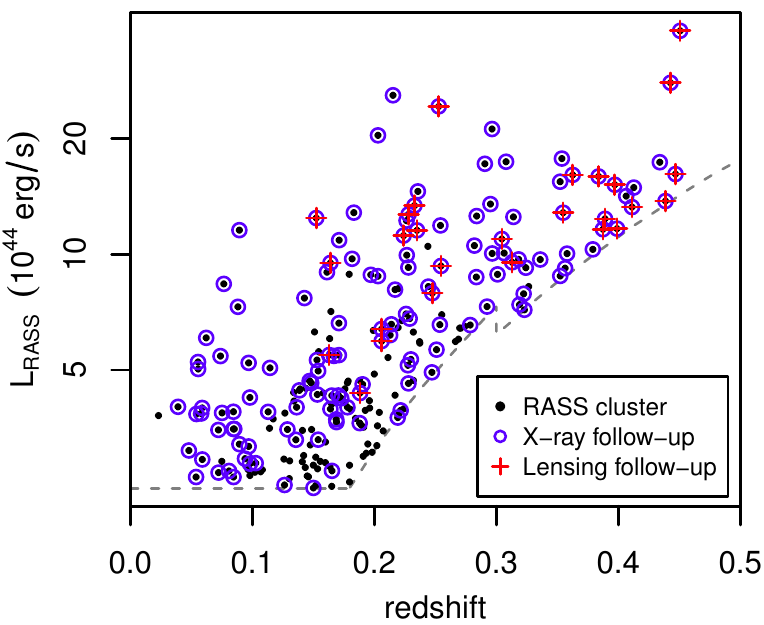}
  \caption{
    Redshift-luminosity distribution of RASS clusters in our data set. The gray, dashed line shows the composite flux limit of the sample (see also Paper~\pfour{}). Circles and crosses respectively indicate clusters for which we employ follow-up X-ray and weak lensing data. The luminosities plotted here are estimated directly from the RASS catalog flux measurements, and are in the rest-frame 0.1--2.4\,keV band.
  }
  \label{fig:selection}
\end{figure}

The data set employed here consists of an X-ray flux limited catalog of clusters culled from the ROSAT All-Sky Survey (RASS; specifically from the BCS, REFLEX and bright MACS samples; \citealt{Ebeling1998MNRAS.301..881E, Ebeling1004.4683, Bohringer0405546}), along with follow-up data providing measurements of weak lensing shear and/or X-ray observables for a subset of the cluster catalog. Altogether, the sample consists of 224 clusters with spectroscopically measured redshifts and X-ray fluxes from the RASS. Of these, 139 have follow-up \Chandra{} and/or ROSAT X-ray data from which we measure temperatures and gas masses (and re-measure X-ray luminosities). We use lensing data to provide mass information for 27 clusters, which are a subset of the 139 with follow-up X-ray data. Figure~\ref{fig:selection} shows the redshift--luminosity distribution of these subsamples. Follow-up X-ray data are available throughout the redshift and X-ray luminosity range spanned by the full sample, while field of view restrictions limit the lensing sub-sample to redshifts $z>0.15$; at these redshifts the lensing data set spans all but the lowest luminosities.

The analysis of the weak lensing data is the subject of Papers~\pone{}--\pthree{}. This section focuses on the \Chandra{} and ROSAT X-ray follow up data used in this work, which updates the analysis of Paper~\pfour{} in two principal ways: the quantity of follow-up data and the version of the \Chandra{} calibration files employed.

\subsection{\Chandra{} data analysis} \label{sec:chandra}

\begin{table*}
  \begin{center}
    \caption{
      Clusters in our sample with follow-up \Chandra{} data. [1] Cluster name; [2--3] J2000 coordinates; [4] clean \Chandra{} exposure time in ks; [5] flag indicating whether weak lensing data are used in our analysis (see Papers~\pone{}--\pthree{}); [6] flag indicating whether a cluster is in the dynamically relaxed sample of \citet{Mantz0909.3098}; [7] \Chandra{} observation IDs used in our analysis. This table continues at the end of the manuscript.
    }
    \label{tab:obsids}
    \vspace{-3ex}
    \begin{tabular}{lccrccc}
      \hline
      Name & RA & Dec & Exp. & WtG & Rel. & OBSIDs \\
      \hline
      Abell~1068           &  10:40:44.596  &  +39:57:11.05    &  28.5   &  ~        &  ~        &  1652,13597                                                                                                                                                                                                                                                                                                                                                                                            \\
Abell~1132           &  10:58:26.136  &  +56:47:43.04    &  8.7    &  ~        &  ~        &  13376                                                                                                                                                                                                                                                                                                                                                                                                 \\
Abell~115            &  00:55:50.351  &  +26:24:35.18    &  306.0  &  ~        &  ~        &  3233,13458,13459,15578,15581                                                                                                                                                                                                                                                                                                                                                                          \\
Abell~1201           &  11:12:54.325  &  +13:26:06.30    &  66.7   &  ~        &  ~        &  4216,7697,9616                                                                                                                                                                                                                                                                                                                                                                                        \\
Abell~1204           &  11:13:20.499  &  +17:35:40.20    &  20.5   &  ~        &  ~        &  2205                                                                                                                                                                                                                                                                                                                                                                                                  \\
Abell~1246           &  11:23:57.744  &  +21:28:51.70    &  4.7    &  ~        &  ~        &  11770                                                                                                                                                                                                                                                                                                                                                                                                 \\
Abell~1413           &  11:55:17.930  &  +23:24:18.10    &  109.3  &  ~        &  ~        &  1661,5002,5003,7696                                                                                                                                                                                                                                                                                                                                                                                   \\
Abell~1423           &  11:57:17.203  &  +33:36:40.24    &  33.0   &  ~        &  ~        &  538,11724                                                                                                                                                                                                                                                                                                                                                                                             \\
Abell~1553           &  12:30:47.457  &  +10:33:10.96    &  11.8   &  ~        &  ~        &  12254                                                                                                                                                                                                                                                                                                                                                                                                 \\
Abell~1650           &  12:58:41.483  &  $-$01:45:44.13  &  203.6  &  ~        &  ~        &  4178,5822,5823,6356,6357,6358,7242,7691                                                                                                                                                                                                                                                                                                                                                               \\
Abell~1651           &  12:59:22.304  &  $-$04:11:46.84  &  9.1    &  ~        &  ~        &  4185                                                                                                                                                                                                                                                                                                                                                                                                  \\
Abell~1664           &  13:03:42.468  &  $-$24:14:43.79  &  46.4   &  ~        &  ~        &  1648,7901                                                                                                                                                                                                                                                                                                                                                                                             \\
Abell~1682           &  13:06:50.086  &  +46:33:28.50    &  17.8   &  ~        &  ~        &  11725                                                                                                                                                                                                                                                                                                                                                                                                 \\
Abell~1689           &  13:11:29.586  &  $-$01:20:29.65  &  178.9  &  ~        &  ~        &  540,1663,5004,6930,7289,7701                                                                                                                                                                                                                                                                                                                                                                          \\
Abell~1763           &  13:35:17.411  &  +41:00:03.85    &  16.0   &  ~        &  ~        &  3591                                                                                                                                                                                                                                                                                                                                                                                                  \\
Abell~1795           &  13:48:52.594  &  +26:35:28.75    &  847.0  &  ~        &  ~        & 493,3666,5286,5287,5288,5289,5290,6159,6160,6161,6162,6163,\\ & & & & & & 10432,10433,10898,10899,10900,10901,12026,12027,12028,12029,\\ & & & & & & 13106,13107,13108,13109,13110,13111,13112,13113,13413,13414,\\ & & & & & & 13415,13416,13417,14268,14269,14270,14271,14272,14273,14274,\\ & & & & & & 14275,15485,15486,15487,15488,15489,15490,15491,15492,16432,\\ & & & & & & 16433,16434,16435,16436,16437,16438,16439,16465,16466,16467,\\ & & & & & & 16468,16469,16470,16471,16472  \\
Abell~1835           &  14:01:01.985  &  +02:52:41.93    &  183.6  &  $\surd$  &  $\surd$  &  496,6880,6881,7370                                                                                                                                                                                                                                                                                                                                                                                    \\

      \hline
   \end{tabular}
  \end{center}
\end{table*} 

The list of \Chandra{} OBSIDs included in our analysis, comprising a total of 9.1\,Ms of clean exposure time for 139 clusters, appears in Table~\ref{tab:obsids}. Our procedure for reducing these data is documented in \citet[][hereafter \morph]{Mantz1502.06020}, with the exception that this work employs a more recent version of the \Chandra{} calibration files, corresponding to {\sc ciao}\footnote{\url{http://cxc.harvard.edu/ciao/}} version 4.6.1 and {\sc caldb}\footnote{\url{http://cxc.harvard.edu/caldb/}} version 4.6.2. In brief, the raw data were reprocessed to produce level 2 event files, and were filtered to eliminate periods of high background. For each observation, a corresponding quiescent background data set was produced using the \Chandra{} blank-sky data,\footnote{\url{http://cxc.cfa.harvard.edu/ciao/threads/acisbackground/}} rescaled according to the measured count rate in the 9.5--12\,keV band. Each cluster field was tested for the presence of a soft Galactic foreground, which is absent from the blank-sky data sets, as described in \citet{Mantz1402.6212}. When present, this foreground component was constrained simultaneously with the cluster model in our subsequent spectral analysis.

We define the cluster center as the ``global center'' produced by the morphology analysis of \morph{}, which is essentially the median photon position in a cluster image after masking point sources. These centers generally agree well with centers defined by iterative centroiding (as in, e.g., \citealt{Vikhlinin0805.2207} and \citealt{Mantz0909.3099}, hereafter \maerd{}), and are straightforward to compute automatically.

We next identified a series of annuli to use for spectral analysis, and extracted spectra following the procedure detailed in \citet{Mantz1402.6212}. For each cluster, background-subtracted, flat-fielded surface brightness profiles were extracted in two energy bands: 0.6--2.0\keV{} and 4.0--7.0\keV{}. The soft-band profiles were used to identify annuli which provide a good sampling of the shape of the surface brightness profile without being dominated by Poisson fluctuations, and with the outermost annulus still containing a clear cluster signal above the background. The hard-band surface brightness profiles were similarly used to define outermost radii where there was clear cluster signal at energies $>4\keV$, a requirement for robustly measuring the temperatures of hot clusters, such as those in our sample. Given the steep drop in \Chandra{}'s effective area at these energies, the outer radius for temperature measurements for a given cluster is smaller than the outer radius where emission is detected in the soft band. For each of these annuli, we generated spectra, response matrices and corresponding blank-sky background spectra, binning the source spectra to have at least one count per channel.

Our spectral analysis was carried out using {\sc xspec}\footnote{\url{http://heasarc.gsfc.nasa.gov/docs/xanadu/xspec/}} (version 12.8.2). Thermal emission from hot, optically thin gas in the clusters, and the local Galactic halo, was modeled as a sum of bremsstrahlung continuum and line emission components, evaluated using the {\sc apec} plasma model (ATOMDB version 2.0.1). Relative metal abundances were fixed to the solar ratios of \citet{Asplund0909.0948}, with the overall metallicity allowed to vary. Photoelectric absorption by Galactic gas was accounted for using the {\sc phabs} model, employing the cross sections of \citet{Balucinska1992ApJ...400..699B}. For each cluster field, the equivalent absorbing hydrogen column densities, \NH{}, were fixed to the values from the HI survey of \citet{Kalberla0504140}, except for cases where the published values are $>10^{21}\cm^{-2}$. The likelihood of spectral models was evaluated using the \citet{Cash1979ApJ...228..939} statistic, as modified by \citet[][the $C$-statistic]{Arnaud1996ASPC..101...17A}. Confidence regions were determined by Markov Chain Monte Carlo (MCMC) explorations of the relevant parameter spaces.

\begin{figure*}
  \centering
  \includegraphics[scale=\figscale]{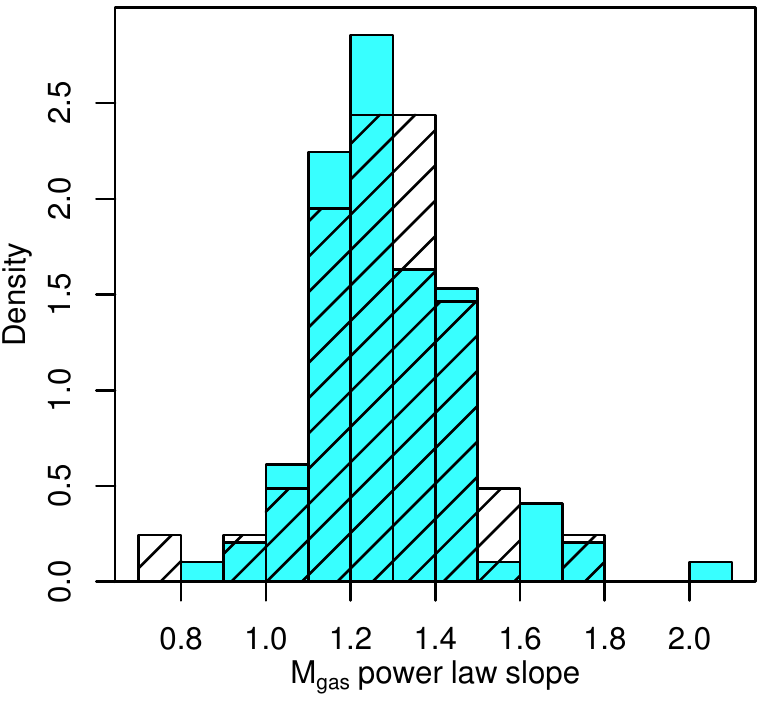}
  \hspace{1cm}
  \includegraphics[scale=\figscale]{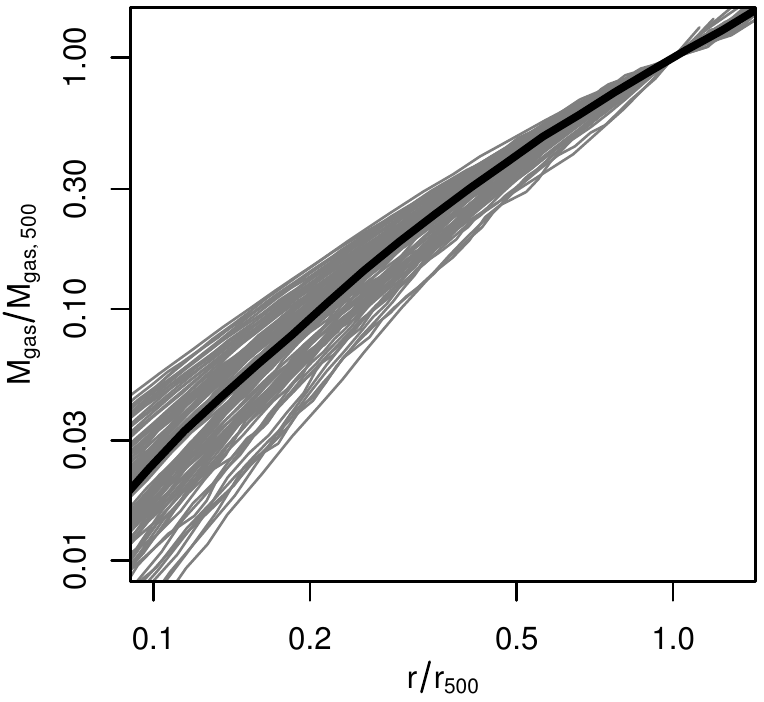}
  \caption{
    Left: histogram of $\Mgas(r)$ power-law slopes at $r_{500}$ from clusters where the measured profiles extend beyond $r_{500}$ (blue shaded), and slopes used to extrapolate profiles that do not reach $r_{500}$ (hatched). Extrapolations are typically no more than 10 per cent in radius.
    Right: scaled gas mass profiles for individual clusters (gray) and the median profile (thick, black line).
  }
  \label{fig:Mgases}
\end{figure*}
 
Our analysis of each cluster is a two-stage process: first a projected analysis to provide an initial estimate of $r_{500}$, and second a deprojection from which mass proxies will be extracted in Section~\ref{sec:proxies}. In the first step, the cluster emission in each annulus was modeled as a redshifted thermal component with Galactic absorption, with independent normalizations in each annulus but linked temperatures and metallicities. For this initial analysis, the temperatures and metallicities were fitted only at radii $>150$\,kpc (comfortably excluding any cool cores). From these fits, we obtained refined estimates of the foreground model parameters (where applicable) and accurate measurements of the (possibly foreground-subtracted) surface brightness profiles from the normalizations of the cluster components in each of the annuli. The best-fitting surface brightness profiles were then geometrically deprojected, accounting for projected emission from a beta-model continuation of the surface brightness at radii larger than the extracted profiles.\footnote{This is an iterative procedure whereby a beta model is fit to the tail of the density profile, the profile is corrected for projected emission, the beta model is re-fit to the corrected densities, and so on until convergence. Typically, only the outermost measured density point is affected at more than the few per cent level by this correction.} These were converted into gas mass profiles, assuming our reference cosmological model and a canonical value of $\mu=0.61\mproton$ for the mean molecular mass of the intracluster medium (ICM). Based on these gas mass profiles, we produced an initial estimate of $r_{500}$ for each cluster by solving the implicit equation
\begin{equation} \label{eq:rDelta}
  M(r_{500}) = \frac{\Mgas(r_{500})}{\fgas(r_{500})} = \frac{4\pi}{3} \, 500\rhocr(z) r_{500}^3,
\end{equation}
using a reference value of the gas mass fraction, $\fgas(r_{500}) = 0.11$.\footnote{As discussed further in Section~\ref{sec:followup}, the reference $\fgas$ used at this stage need not be precisely correct (this particular value was used for historical reasons). We only require an $r_{500}$ estimate which falls relatively close to the true $r_{500}$ in terms of the slope of the $\Mgas$ profile. Given the weak dependence on $\fgas$, $r_{500}\propto\fgas^{-1/3}$, the difference between 0.11 and the value of $\sim0.125$ that ultimately results from our analysis is immaterial for this purpose.}

In the second analysis step, the data for each cluster were fitted with a non-parametric model for the deprojected, spherically symmetric ICM density and temperature profiles (the {\sc projct} model in {\sc xspec}). In this model, the cluster atmosphere is described as a set of concentric, spherical shells, with radii corresponding to the set of annuli from which spectra were extracted. While the density in each shell was free, the model included only two free temperatures, corresponding to the cluster volumes at radii $<0.15\,r_{500}$ and $>0.15\,r_{500}$, for the $r_{500}$ estimate produced in the previous step. Metallicities were linked in the same way. The temperature measured in the outer radial bin is essentially identical to what we would obtain from a typical analysis of the spectrum in projection, since projected emission from larger radii is not accounted for. The advantage of this approach over the simpler strategy of measuring temperature in projection from a single spectrum and (separately) gas mass from a surface brightness profile is that it allows the measurement covariance between the projected ``center-excised'' temperature, the projected luminosity, and the spherically integrated (deprojected) gas mass to be fully captured (see Section~\ref{sec:proxies}). The gas density profiles resulting from these fits were again corrected for projected emission from radii outside the modelled regions, assuming a beta-model continuation of the surface brightness.

\subsection{ROSAT data analysis} \label{sec:rosat}

For the most nearby clusters in our sample, the \Chandra{} field of view precludes measurements of the cluster surface brightness at radii close to $r_{500}$. For these clusters, we take advantage of larger field-of-view ROSAT PSPC data. The reduction and analysis of these data are described in \maerd{}, and we use the deprojected gas mass profiles derived in that work. To account for any differences in flux calibration between \Chandra{} and ROSAT, we rescale each ROSAT gas mass profile to match the \Chandra{} profile for the same cluster at radii where they overlap, excluding small radii where the larger point spread function of ROSAT is significant. For the 21 clusters where we use ROSAT profiles, the correction factor from ROSAT to \Chandra{} was $1.11\pm0.11$.

\begin{figure*}
  \centering
  \includegraphics[scale=\figscale]{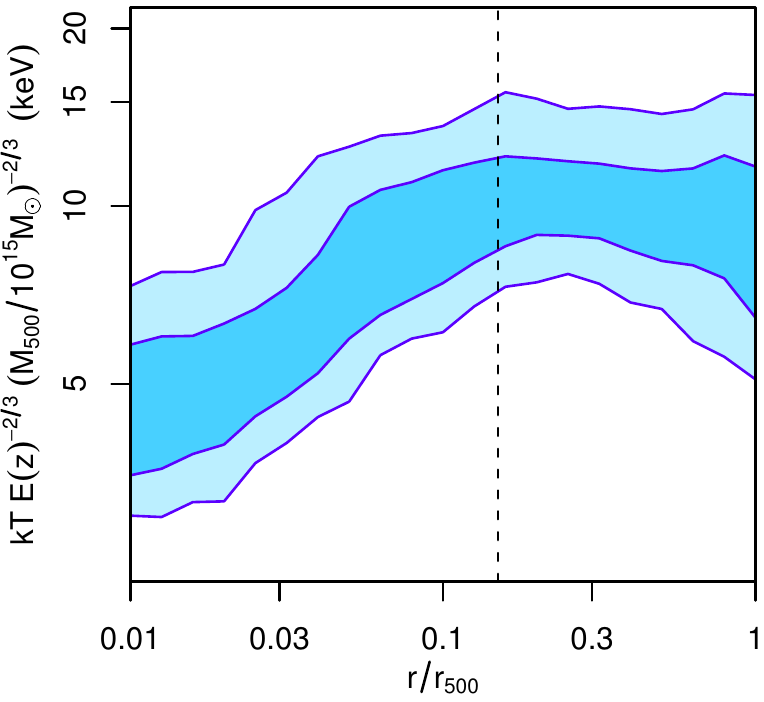}
  \hspace{1cm}
  \includegraphics[scale=\figscale]{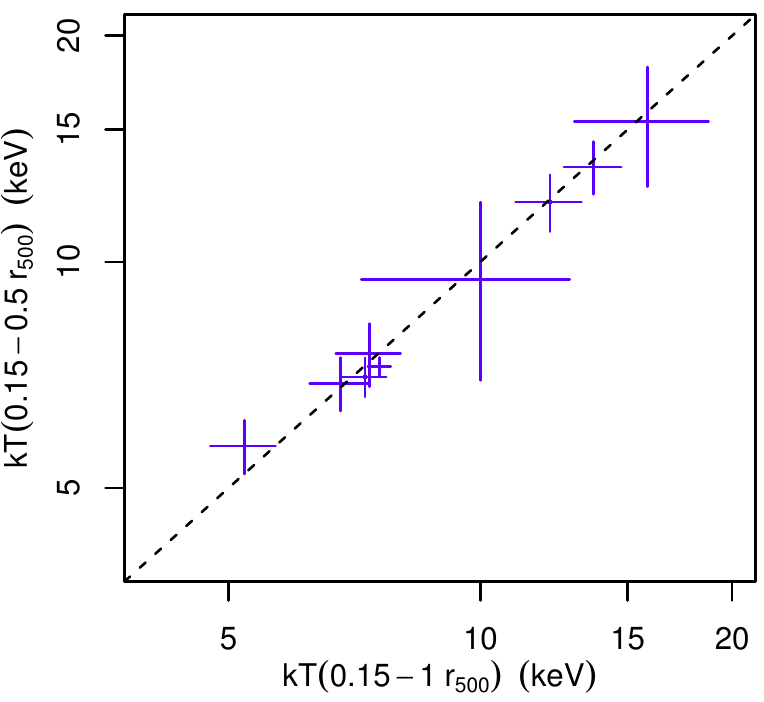}
  \caption{
    Left: ensemble temperature profile, accounting for self-similar scaling, from the relaxed cluster analysis of \citet{Mantz1509.01322}. The vertical, dashed line shows the inner radius of the nominal aperture (0.15--1\,$r_{500}$) for average temperature measurements in this work, which effectively excludes the central decrement associated with cool cores.
    Right: comparison of temperatures measured in a 0.15--0.5\,$r_{500}$ aperture with those measured at radii of 0.15--1\,$r_{500}$, for the 9 clusters where the hard-band signal-to-noise is sufficient to provide information out to $>0.9\,r_{500}$. Differences between the two measurements are at the per cent level, in agreement with the aperture corrections predicted from the scaled ensemble profile.
  }
  \label{fig:kTap}
\end{figure*}
  
\subsection{Obtaining mass proxies} \label{sec:proxies}

The output of the analysis above is, for each cluster, a list of monte-carlo realizations of the ICM density profile (from either \Chandra{} or ROSAT) and the average temperature measured in a center-excised aperture (from \Chandra{}). With a few more steps, these can be converted into the quantities that will be used in our scaling relation analysis, namely (1) the intrinsic 0.1--2.4\,keV (rest-frame) luminosity projected within $r_{500}$, (2) the average, emission-weighted, projected temperature measured in an annulus spanning 0.15--1\,$r_{500}$, and (3) the spherically integrated gas mass within $r_{500}$. The following procedure was performed for every monte-carlo realization of each cluster, producing distributions of these mass proxies for each cluster.

For each realization, new values of $r_{500}$ and $\Mgas(r_{500})$ are derived from the gas mass profile according to Equation~\ref{eq:rDelta}; however, this time we marginalize over an expected bias of $3\pm6$ per cent in the amplitude of the $\Mgas(r)$ profile at $r_{500}$, due to asphericity and projection effects, based on the hydrodynamical simulations of \citet{Nagai0609247}. Where necessary in order to satisfy Equation~\ref{eq:rDelta}, we fit a power law to the outermost portion of the measured $\Mgas$ profile and extrapolate it (typically by no more than 10 per cent in radius). The distribution of power-law slopes used in this step is in excellent agreement with the slopes at $r_{500}$ measured from the subsample of clusters whose profiles extend beyond $r_{500}$ (left panel of Figure~\ref{fig:Mgases}), indicating that such modest extrapolations do not introduce a systematic bias in our $\Mgas$ estimates. This is a reflection of the strong similarity of cluster gas density profiles at $\sim r_{500}$ extensively commented on in the literature (e.g.\ \citealt{Croston0801.3430, Pratt0809.3784, Mantz1509.01322}), and demonstrated in the $\Mgas$ profiles from this analysis in the right panel of Figure~\ref{fig:Mgases}.

Next, we derive the unabsorbed, projected, intrinsic luminosity profile of each cluster realization in the desired energy band of 0.1--2.4\,keV, accounting for the temperature- and redshift-dependent K-corrections. Values of the projected luminosity within the estimated value of $r_{500}$ are straightforwardly extracted.

The final step is to apply an aperture correction to the center-excised, emission-weighted temperatures. For consistency with previous work, we would like these temperatures to correspond to radii of 0.15--1\,$r_{500}$. In practice, the outer radii for temperature measurement identified in Section~\ref{sec:chandra} are almost always $<r_{500}$, while the inner radii may differ slightly from $0.15\,r_{500}$ simply due to the initial choice boundaries between annuli. To account for the small aperture dependence of these temperature measurements, we use the ensemble scaled temperature profile of relaxed clusters measured by \citet{Mantz1509.01322}, shown as a function of $r/r_{500}$ in the left panel of Figure~\ref{fig:kTap}. By design, the temperature gradient at small radii ubiquitous to relaxed clusters (but not to the cluster population at large) does not impact the temperature profile at the target radii of 0.15--1\,$r_{500}$, so this relaxed cluster profile template can be fairly used to correct for small differences in the measurement aperture, particularly where the outer radius is concerned. Based on this temperature template, accounting for emission weighting according to the corresponding density profile template, we calculate that aperture corrections for our clusters are at the per cent level, even for cases where the outer measurement radii are $\sim 0.3\, r_{500}$. As a sanity check, we repeated our analysis for the 9 clusters where the temperature measurement region reached $>0.9\,r_{500}$, this time using an outer radius for temperature measurement half as large. The results are shown in the right panel of Figure~\ref{fig:kTap}; as expected, the two measurements agree at the per cent level on average.\footnote{Note that our approach fundamentally differs from that used in some earlier works (\citealt{ Vikhlinin0805.2207}; \maerd{}) in which the temperature measured in an aperture extending to $r_{500}$ was treated as truth, regardless of the actual signal-to-noise or background uncertainties. Comparisons of half- to full-aperture measurements in those works, necessitated by Chandra's limited field of view for low-$z$ clusters, produced typical corrections of $\sim10$ per cent.}

\begin{table*}
  \begin{center}
    \caption{
      Mass proxies from X-ray and weak lensing data for clusters in our data set, where available. [1] Cluster name; [2] redshift; [3] scale radius, $r_{500}$, defining the apertures within which other quantities are measured; [4] spherically integrated gas mass within $r_{500}$; [5] projected temperature measured in an aperture spanning projected radii of 0.15--1\,$r_{500}$; [6] 0.1--2.4\,keV intrinsic, rest-frame luminosity projected within $r_{500}$; spherically integrated gravitating mass within $r_{500}$ determined from weak lensing data (see Paper~\pthree{}). Measured redshifts and temperatures have no direct dependence on cosmological assumptions; other quantities are derived assuming a flat \LCDM{} model with $h=0.7$ and $\Omegam=0.3$. In addition, a gas mass fraction of $\fgas(r_{500})=0.125$ is used to determine the characteristic radii, $r_{500}$. The listed values are thus referenced to the particular cosmological model and gas mass fraction above, a fact which must be accounted for when exploring cosmological and scaling relation models, as described in the text. This table continues at the end of the manuscript.
    }
    \label{tab:proxies}
    \vspace{1ex}
    \begin{tabular}{lcr@{$\pm$}lr@{$\pm$}lr@{$\pm$}lr@{$\pm$}lc}
      \hline
      Name & $z$ & \multicolumn{2}{c}{$r_{500}$} & \multicolumn{2}{c}{$\Mgas$} & \multicolumn{2}{c}{$kT$} & \multicolumn{2}{c}{$L$} & $\Mlens$ \\
      &  & \multicolumn{2}{c}{(Mpc)} & \multicolumn{2}{c}{$(10^{14}\Msun)$} & \multicolumn{2}{c}{(keV)} & \multicolumn{2}{c}{$(10^{44}\erg\second^{-1})$} & $(10^{15}\Msun)$ \\
      \hline
      Abell~3571           &  0.039  &     1.25  &         0.04  &     0.72  &         0.08  &   6.26   &       0.27  &   4.03   &       0.06  &    ---                     \\
Abell~3558           &  0.048  &     1.29  &         0.04  &     0.79  &         0.08  &   6.09   &       0.27  &   4.32   &       0.15  &    ---                     \\
Abell~754            &  0.054  &     1.21  &         0.05  &     0.66  &         0.07  &   9.40   &       0.24  &   2.75   &       0.08  &    ---                     \\
Hydra                &  0.054  &     0.94  &         0.03  &     0.31  &         0.03  &   4.13   &       0.09  &   3.28   &       0.03  &    ---                     \\
Abell~3667           &  0.056  &     1.40  &         0.07  &     1.03  &         0.15  &   6.56   &       0.13  &   4.64   &       0.26  &    ---                     \\
Abell~85             &  0.056  &     1.29  &         0.04  &     0.80  &         0.08  &   7.79   &       0.19  &   5.44   &       0.16  &    ---                     \\
Abell~2256           &  0.058  &     1.47  &         0.05  &     1.19  &         0.13  &   7.12   &       0.24  &   7.71   &       0.29  &    ---                     \\
Abell~3158           &  0.059  &     1.15  &         0.04  &     0.58  &         0.07  &   5.46   &       0.13  &   2.96   &       0.04  &    ---                     \\
Abell~3266           &  0.059  &     1.49  &         0.06  &     1.23  &         0.15  &   10.00  &       0.31  &   6.10   &       0.22  &    ---                     \\
Abell~1795           &  0.063  &     1.21  &         0.04  &     0.66  &         0.06  &   7.47   &       0.16  &   6.26   &       0.06  &    ---                     \\
Abell~2065           &  0.072  &     1.16  &         0.04  &     0.60  &         0.06  &   6.58   &       0.21  &   3.19   &       0.05  &    ---                     \\
Abell~399            &  0.072  &     1.31  &         0.05  &     0.86  &         0.10  &   7.09   &       0.27  &   3.11   &       0.05  &    ---                     \\
Abell~401            &  0.074  &     1.62  &         0.05  &     1.63  &         0.15  &   9.67   &       0.22  &   9.79   &       0.36  &    ---                     \\
Abell~3112           &  0.075  &     1.14  &         0.05  &     0.56  &         0.07  &   6.08   &       0.16  &   5.75   &       0.22  &    ---                     \\
Abell~2029           &  0.078  &     1.44  &         0.05  &     1.16  &         0.13  &   9.76   &       0.24  &   12.29  &       0.36  &    ---                     \\
Abell~2255           &  0.081  &     1.26  &         0.05  &     0.77  &         0.10  &   7.17   &       0.26  &   3.43   &       0.22  &    ---                     \\
Abell~1650           &  0.084  &     1.15  &         0.04  &     0.58  &         0.06  &   6.29   &       0.14  &   4.05   &       0.04  &    ---                     \\
Abell~1651           &  0.084  &     1.25  &         0.05  &     0.75  &         0.09  &   7.63   &       0.40  &   5.54   &       0.29  &    ---                     \\
Abell~2420           &  0.085  &     1.15  &         0.04  &     0.58  &         0.06  &   6.98   &       0.43  &   3.06   &       0.07  &    ---                     \\

      \hline
   \end{tabular}
  \end{center}
\end{table*}

 Table~\ref{tab:proxies} contains mass proxy measurements obtained using the procedure described above for our reference cosmology, as well as weak lensing masses estimated within the same radii. To make them as widely useful as possible, the tabulated values are derived using the reference $\fgas(r_{500})=0.125$, the best fit resulting from our later analysis (Section~\ref{sec:baseline}), in the determination of $r_{500}$. Note that the procedure for fitting the cosmology+scaling relations model properly accounts for the assumed reference when comparing predictions to the measured values (Section~\ref{sec:followup}).

\subsection{External data sets}

While the primary data for constraining cluster scaling relations are those discussed above, we employ additional data sets to help constrain the background cosmological parameters. Our ``clusters only'' analysis in Section~\ref{sec:results} uses the catalogs and follow-up data described above in combination with cluster gas-mass fraction data at radii $\sim r_{2500}$ \citep{Mantz1402.6212}, which provide additional constraints on the expansion history of the Universe, and Gaussian priors on the Hubble parameter ($h=0.738\pm0.024$; \citealt{Riess1103.2976}) and the cosmic baryon density ($100\,\Omegab h^2=2.202\pm0.045$; \citealt{Cooke1308.3240}). Our ``All data'' results do not use the priors on $h$ or $\Omegab h^2$, but instead incorporate all of the cluster data along with cosmic microwave background (CMB), type Ia supernova and baryon acoustic oscillation (BAO) data. The CMB data set includes \Planck{} 1-year data supplemented by WMAP polarization measurements \citep{Planck1303.5075}), and high-multipole data from the Atacama Cosmology Telescope (ACT; \citealt{Das1301.1037}) and the South Pole Telescope (SPT; \citealt{Keisler1105.3182, Reichardt1111.0932, Story1210.7231}. The supernova data set employed is the Union 2.1 compilation of type Ia supernovae \citep{Suzuki1105.3470}, and the BAO data set combines results from the 6-degree Field Galaxy Survey (6dF; $z=0.106$; \citealt{Beutler1106.3366}) and the Sloan Digital Sky Survey (SDSS, $z=0.35$ and $0.57$; \citealt{Padmanabhan1202.0090, Anderson1303.4666}). See Paper~\pfour{} and \citet{Mantz1402.6212} for more details of how we use these data. For this work, the purpose of including the non-cluster data is to constrain the cosmological model as well as possible (while still marginalizing over its remaining uncertainty), so that the cluster data can most effectively probe the scaling relations. Note, however, that the difference between the ``clusters only'' and ``all data'' results on the scaling relations are small compared to the final uncertainties. We therefore do not expect that, e.g., using the final \Planck{} CMB data set would change our results noticeably.

\section{Model} \label{sec:model}

\vspace{-0.1cm}
The model employed here, which includes cosmology, scaling relations, and sampling distributions for both the survey and follow-up measurements, is described in \mare{} and Paper~\pfour{}. Here we review the aspects most relevant for this work, namely the scaling relation model and the sampling distribution model for the follow-up X-ray observations. In terms of an overall likelihood of the data set given a model, the context for the expressions below is the discussion in Section 3.3 of Paper~\pfour{}; we particularly refer the reader there for details of the cosmological modeling and the sampling distribution for the weak lensing data. Throughout this work, we assume a flat \LCDM{} cosmological model.

\subsection{Scaling relations} \label{sec:scalingmodel}

We define the logarithmic mass within $r_{500}$ as\footnote{We note a minor correction to the notation used in Paper~\pfour{}, which affects Equations~\ref{eq:scalingmass}, \ref{eq:MLTdefs}, \ref{eq:ltmgas-scaling} and \ref{eq:expandedmodel}. Namely, factors of $E(z)$ should not have been bundled in to the definitions of $m$, $\mgas$ and $\mlens$. The definitions used here are the correct expression of a model with self-similar evolution, and correspond to the actual implementation used in both Paper~\pfour{} and this work. We note, however, that because the slopes of the $\mgas$ and $\mlens$ scaling relations are very close to unity, this would have made a correspondingly small difference in any case.}
\begin{eqnarray}
  \label{eq:scalingmass}
  m &=& \ln\left(\frac{M_{500}}{10^{15}\Msun}\right).
\end{eqnarray}
The properties whose scaling relations are to be modeled are
\begin{eqnarray}
  \label{eq:MLTdefs}
  \ell &=& \ln\left(\frac{L_{500}}{E(z)10^{44}\erg\second^{-1}}\right),  \\
  t &=& \ln\left(\frac{kT_{500}}{\keV}\right), \nonumber\\
  \mgas &=& \ln\left(\frac{\Mgas{}_{,500}}{10^{15}\Msun}\right), \nonumber\\
  \mlens &=& \ln\left(\frac{\Mlens{}_{,500}}{10^{15}\Msun}\right), \nonumber
\end{eqnarray}
where $E(z)=H(z)/H_0$, $L_{500}$ is cluster rest-frame luminosity in the 0.1--2.4\,keV band, $kT_{500}$ is the emission-weighted temperature measured in an annulus spanning 0.15--1\,$r_{500}$, $\Mgas{}_{,500}$ is the gas mass within $r_{500}$, and $\Mlens{}_{,500}$ is the spherical mass estimate from lensing corresponding to an idealized shear profile without statistical noise (but including the effects of projected structure). For convenience, we also define
\begin{eqnarray}
  \varepsilon &=& \ln\left[ E(z) \right].
\end{eqnarray} 

Each of the quantities in Equations~\ref{eq:scalingmass}--\ref{eq:MLTdefs} is idealized: $m$ refers to the true, unknowable mass of a cluster, while the components of $\bmath{y} \equiv (\ell,t,\mgas,\mlens)$ represent intrinsic properties that will differ from observed values due to measurement error. Note that the values of $m$ and $\bmath{y}$ for each cluster are nuisance parameters of the model. While $\bmath{y}$ can be directly constrained by the survey flux and/or follow-up observations, $m$ is only ever constrained indirectly by observations of $\bmath{y}$ in conjunction with the scaling relation model and the prior probability distribution of $m$ (the mass function; see \mare{} and Paper~\pfour{}).

The baseline scaling relation model considered in this work consists of a power-law describing the mean of each response variable, with evolutionary terms corresponding to the self-similar model \citep{Kaiser1986MNRAS.222..323K}, and a log-normal distribution accounting for intrinsic scatter at fixed mass and redshift. The mean scalings are
\begin{eqnarray} \label{eq:ltmgas-scaling}
  \expectation{\ell} &=& \beta_{0,\ell} + \beta_{1,\ell}(\varepsilon+m), \\
  \expectation{t} &=& \beta_{0,t} + \beta_{1,t}(\varepsilon+m), \nonumber\\
  \expectation{\mgas} &=& \beta_{0,\mgas} + \beta_{1,\mgas} \, m, \nonumber\\
  \expectation{\mlens} &=& \beta_{0,\mathrm{lens}} + \beta_{1,\mlens} \, m, \nonumber
\end{eqnarray}
Note that $\beta_{0,\mgas}$ is $\ln\,\fgas(r_{500})$ for an $M_{500}=10^{15}\Msun$ cluster. The corresponding likelihood function is
\begin{equation} \label{eq:scaling}
  P(\bmath{y}|m) \propto |\mathbf{\Sigma}|^{-1/2} \exp\left( -\frac{1}{2} \bmath{\eta}^\mathrm{t} \mathbf{\Sigma}^{-1} \bmath{\eta} \right) ,
\end{equation}
where $\mathbf{\Sigma}$ is a covariance matrix and $\bmath{\eta}=\bmath{y}-\bmath{\expectation{y}}$. 

The normalizations ($\bmath{\beta_0}$) and slopes ($\bmath{\beta_1}$) of the nominal scaling relations, and the elements of $\mathbf{\Sigma}$, together comprise 18 parameters. Several of these can be either fixed or informed by priors in order to reduce the complexity of the model (and thereby the computational cost of evaluating the likelihood). In the mean scaling of $\mlens$, we fix the slope, $\beta_{1,\mlens}$, to unity and marginalize over a Gaussian prior on the normalization, $\beta_{0,\mlens} = -0.01 \pm 0.07$, encoding the mean bias and systematic uncertainty of the WtG ``color-cut'' lensing analysis (see Paper~\pthree{}). Uninformative priors are used for the normalizations and slopes of the nominal $\ell$, $t$ and $\mgas$ scalings. In principle, the absolute value and evolution of $\fgas(r_{500})$, the $\mgas$--$m$ normalization, carry cosmological information \citep{Sasaki9611033, Pen9610090, Allen0205007, Allen0405340, Allen0706.0033, Allen1103.4829, Ettori0211335, Ettori0904.2740, Battaglia1209.4082, Planelles1209.5058, Mantz1402.6212}. In practice, however, the low precision of our individual lensing mass constraints makes this information subordinate to the constraining power available from cluster counts and from the more precise $\fgas$ measurements possible at $\sim r_{2500}$ \citep{Mantz1402.6212}. We nevertheless marginalize over a prior for the evolution in the normalization of the \mgas{}--$m$ relation throughout our analysis, $\fgas(r_{500},z) = \fgas(r_{500},z=0)(1+\alpha_f z)$, with $-0.05<\alpha_f<0.05$ (see further discussion in Paper~\pfour{}).

We expand the intrinsic covariance matrix as
\begin{equation} \label{eq:cov}
  \bmath{\Sigma} = \left(\begin{array}{cccc}
      \sigma^2_\ell & \rho_{\ell t}\sigma_\ell\sigma_t & 0 & 0 \\
      \rho_{\ell t}\sigma_\ell\sigma_t & \sigma^2_t & 0 & 0 \\
      0 & 0 & \sigma^2_{\mgas} & 0 \\
      0 & 0 & 0 & \sigma^2_{\mlens}
    \end{array}\right),
\end{equation}
in which most off-diagonal elements have been fixed to zero. While this simplification is required for computational reasons, it is also well motivated according to our best understanding of the observables involved. The marginal scatter in X-ray luminosity at fixed mass is dominated by the presence or absence of compact, bright cores found at the centers of some clusters (e.g.\ \citealt{Fabian1994MNRAS.267..779F, Allen9802218, Markevitch9802059, Peres9805122}). This luminosity scatter is both large ($\sim40$ per cent) and physically different in origin from the scatters in \Mgas{} and \Mlens{}, both of which are most sensitive to larger spatial scales ($\sim r_{500}$ compared with $\ltsim 0.05\,r_{500}$). Covariances among the temperature, gas mass and lensing mass measured for a given cluster, due to e.g.\ asphericity or dynamical state, are similarly thought to be small (see e.g.\ calculations by \citealt{Gavazzi0503696} and \citealt{Buote2012MNRAS.421.1399B}, and hydrodynamical simulations by \citealt{Stanek0910.1599}).\footnote{Relatively weak empirical constraints (not accounting for the mass function and selection biases) on some of these terms are presented by \citet{Maughan1212.0858} and \citet{Mantz1509.01322}.} Thus, while constraining these off-diagonal terms remains an interesting avenue for future work, we expect the impact of neglecting them to be small for the present analysis.

Among the remaining parameters, we adopt an informative prior only for the intrinsic scatter in \mgas{}. We require $\sigma_{\mgas} < 0.11$, where 0.11 corresponds to the high end of the confidence interval for the intrinsic scatter of $\fgas(r_{500})$, measured using hydrostatic mass estimates for relaxed clusters \citep{Mantz1509.01322}, and encompasses the expected scatter among all massive clusters from hydrodynamic simulations (e.g.\ \citealt{Battaglia1209.4082}). While Paper~\pfour{} used an informative prior for $\sigma_{\mlens}$, we forgo it here because the data are able to constrain this parameter on their own.

Beyond the baseline scaling relation model described above, we consider in Section~\ref{sec:extended} extensions to the luminosity--mass and temperature--mass relation models. Specifically, we allow departures from self-similar evolution in the normalizations of each relation, of the form
\begin{eqnarray} \label{eq:expandedmodel}
  \expectation{\ell} &=& \beta_{0,\ell} + \beta_{1,\ell}(\varepsilon+m) + \beta_{2,\ell} \, \ln(1+z), \\
  \expectation{t} &=& \beta_{0,t} + \beta_{1,t}(\varepsilon+m) + \beta_{2,t} \, \ln(1+z). \nonumber
\end{eqnarray}
In addition, we consider possible evolution in the luminosity--temperature block of the intrinsic covariance matrix of the form 
\begin{eqnarray} \label{eq:expandedmodelscat}
  \sigma_\ell &\rightarrow& \sigma_\ell (1 + \sigma'_\ell\,z) \\
  \sigma_t &\rightarrow& \sigma_t (1 + \sigma'_t\,z) \nonumber\\
  \rho_{\ell t} &\rightarrow& \rho_{\ell t} (1 + \rho'_{\ell t}\,z). \nonumber
\end{eqnarray}
Lastly, we allow for the possibility of asymmetry in the luminosity--mass intrinsic scatter of the skew-log-normal form (see \citealt{Azzalini4615982, Azzalini2337278, Azzalini2680724}), through the shape parameter, $\lambda_\ell$, where $\lambda_\ell=0$ corresponds to the baseline log-normal model.

\subsection{Follow-up observations} \label{sec:followup}

While there is good physical motivation for defining the scaling relations in terms of $r_{500}$, as above, a consequence is that the corresponding volume of each cluster depends on cosmological parameters (through the critical density) and on the cluster mass, which is a nuisance parameter of the model. In addition, calculations involving cosmological distance are in some cases required to infer the physically interesting quantities appearing in the scaling relations from direct observables (X-ray count rates, spectra, and shear profiles), even within a fixed angular aperture.

For the X-ray observables, the latter effect is simply a scaling by the cosmic distance to a given cluster raised to a power that depends on the observable in question.\footnote{We do not distinguish between angular diameter distances and luminosity distances in this discussion, since extraneous factors of $1+z$ will cancel when we take ratios of the distance calculated in different cosmologies.} Combined with the fact that the ICM density and temperature profiles follow well defined power laws at radii $\sim r_{500}$ (e.g.\ \citealt{Vikhlinin0507092, Allen0706.0033, Croston0801.3430, Mantz1509.01322}), this allows us to take the computationally simple approach of deriving physically meaningful quantities for a cluster assuming a reference cosmological model and mass, and then predicting what would have been measured (under the same assumptions) if the true cosmology and/or mass were different (\mare{}). Here the reference (see Section~\ref{sec:proxies}) consists of a specified flat \LCDM{} cosmology, and a value of $\fgas(r_{500})$ used to solve for $r_{500}$ via Equation~\ref{eq:rDelta}. Given this approach to defining the reference $r_{500}$, and using that $\rhocr(z) \propto H(z)^2$, the ratio of angular distances corresponding to the reference $r_{500}$ and the true $r_{500}$ (according to a given trial model) is
\begin{equation} \label{eq:r500ratio}
  R_\theta = \frac{\theta_{500}^\mathrm{ref}}{\theta_{500}} \approx\left[ \frac{\fgas H(z)^2 d(z)^{1/2}}{\fgas{}_{,\mathrm{ref}} H_\mathrm{ref}(z)^2 d_\mathrm{ref}(z)^{1/2}} \right]^{1/(3-\eta_g)},
\end{equation}
where $\eta_g$ is the power-law slope of $\Mgas(r)$ at $r_{500}$ and $d(z)$ is the distance to the cluster. Within a given angular aperture, the gas mass determined from an X-ray observation scales as $d(z)^{5/2}$. When predicting the gas mass that would be measured for a cluster (under the reference assumptions) for a given value of the nuisance parameter $\mgas$, we must therefore scale by the factor
\begin{equation}
  \frac{\Mgas{}_{,\mathrm{ref}}}{\Mgas} \approx \left[\frac{d_\mathrm{ref}(z)}{d(z)}\right]^{5/2} R_\theta^{\eta_g}. 
\end{equation}
The corresponding factor for luminosity measurements is
\begin{equation} \label{eq:Lratio}
  \frac{L_\mathrm{ref}}{L} \approx \left[\frac{d_\mathrm{ref}(z)}{d(z)}\right]^{2} R_\theta^{\eta_L}, 
\end{equation}
where the slope of the projected luminosity profile at $r_{500}$ is $\eta_L \approx 0.1$ on average. A similar expression could be written for temperature; however, given the lack of a fixed-aperture cosmology dependence for temperature and the flatness of cluster temperature profiles over the relevant range (Figure~\ref{fig:kTap}), we neglect such a correction.

Since the three X-ray observables for a given cluster are derived from the same observations, their measurement errors are correlated. This correlation, along with the marginal uncertainty in each quantity, can be measured straightforwardly from the monte carlo samples generated when fitting the data for each cluster in Section~\ref{sec:chandra}. Using a hat to denote measured, as opposed to intrinsic, quantities, we find the measurement correlation coefficients $\mathrm{cor}(\hat{\ell},\hat{t})=-0.29\pm0.14$, $\mathrm{cor}(\hat{\ell},\mysub{\hat{m}}{gas})=0.40\pm0.15$ and $\mathrm{cor}(\hat{t},\mysub{\hat{m}}{gas})=0.01\pm0.04$ across the entire sample. Incorporating these measurement covariances and Equations~\ref{eq:r500ratio}--\ref{eq:Lratio}, the likelihood associated with the follow-up X-ray data for a cluster, $P(\hat{\ell},\hat{t},\hat{m}_\mathrm{gas}|z,\ell,t,\mgas)$, is approximated as a multivariate Gaussian.

\section{X-ray Scaling Relations} \label{sec:results}

This section presents our constraints on the cluster scaling relations. Results for the baseline model are covered in Section~\ref{sec:baseline} and compared with the literature in Section~\ref{sec:comparisons}, while extensions to the model are discussed in Section~\ref{sec:extended}.

\subsection{Baseline model} \label{sec:baseline}

Table~\ref{tab:baseline} presents the marginalized constraints on parameters of the scaling relations model in three cases: (1) using the ``clusters only'' data set and marginalizing over cosmological parameters, (2) using ``all data'' and marginalizing over cosmological parameters, and (3) using the ``clusters only'' data and fixing $h=0.7$ and $\Omega_m=0.3$. In the latter case, we still marginalize over the remaining parameters of the flat \LCDM{} cosmological model (in particular $\sigma_8$); this set of results is intended to provide a reference in which degeneracies with the cosmic expansion history do not enter. For the results where the cosmological model is marginalized over, constraints on the scaling relation normalizations are given with the primary cosmological degeneracy (with $h$) explicitly factored out.

\begin{table}
  \begin{center}
    \caption{
      Best-fitting values and marginalized 68.3 per cent confidence intervals for parameters of the baseline scaling relations model. Parameters that are only constrained by their priors, as opposed to the data, are not shown. The first two columns show results obtained using clusters-only data, and using additional data to constrain the background cosmology, respectively. Those results are marginalized over the cosmological parameters of a flat \LCDM{} model. In the last column, the values of $h$ and $\Omegam$ were fixed to 0.7 and 0.3, respectively, but other cosmological parameters were still marginalized over. For the normalization of each scaling relation, we quote the constraints after factoring out the dominant cosmological degeneracy, which is with $h$. Note that the values of $\gamma_\ell$, $\gamma_t$, $\gamma_g$ and $\gamma_y$ that encode this degeneracy depend on which data set is used. For the ``clusters only'' and ``all data'' cases, they are, respectively, $\gamma_\ell=0.69$ and $-4.2$, $\gamma_t=-0.60$ and $-2.4$, $\gamma_g=1.50$ and $-0.19$, and $\gamma_y=0.90$ and $-2.59$. The last four lines show constraints on $\fgas(r_{500})$ and the scaling relation for $\Yx=\Mgas kT$ that are derived from the baseline results.
    }
    \label{tab:baseline}
    \vspace{-2ex}
    \begin{tabular}{l@{ }r@{ $\pm$ }l@{ }r@{ $\pm$ }l@{ }r@{ $\pm$ }l}
      \hline
      Parameter & \multicolumn{2}{c}{Clusters only} & \multicolumn{2}{c}{All data} & \multicolumn{2}{c}{Clusters only} \\
      & \multicolumn{2}{c}{} & \multicolumn{2}{c}{} & \multicolumn{2}{c}{(fixed $h$, $\Omegam$)} \\
      \hline
      $\beta_{0,\ell}+\gamma_\ell\ln(h_{70})$  &  $1.76$         &  $0.16$   &        $1.70$   &        $0.09$   &  $1.65$   &   $0.14$   \\
      $\beta_{1,\ell}$        &  $1.31$         &  $0.06$   &         $1.34$   &        $0.05$   &       $1.35$    &  $0.06$   \\
      $\sigma_\ell$        &  $0.42$         &  $0.03$   &         $0.43$   &        $0.03$   &       $0.43$    &  $0.03$   \\
      $\beta_{0,t}+\gamma_t\ln(h_{70})$  &  $2.19$         &  $0.06$  &        $2.17$   &        $0.04$   &  $2.16$   &   $0.06$   \\
      $\beta_{1,t}$        &  $0.60$         &  $0.04$   &         $0.62$   &        $0.04$   &       $0.63$    &  $0.03$   \\
      $\sigma_t$        &  $0.156$        &  $0.017$  &         $0.160$  &        $0.017$  &       $0.161$   &  $0.019$  \\
      $\rho_{\ell t}$        &  $0.53$         &  $0.10$   &         $0.56$   &        $0.10$  &       $0.54$    &  $0.10$   \\
      $\sigma_{\mlens}$        &  $0.18$         &  $0.06$   &         $0.17$   &        $0.06$   &       $0.17$    &  $0.05$   \\
      $\beta_{0,\mgas}$+$\gamma_g\ln(h_{70})$  &  $-2.05$        &  $0.07$    &        $-2.08$  &        $0.04$  &  $-2.09$  &   $0.06$   \\
      $\beta_{1,\mgas}$            &  $1.004$        &  $0.014$  &         $1.007$  &        $0.012$  &       $1.006$   &  $0.011$  \\
      \hline
      $\fgas(r_{500})\,h_{70}^{\gamma_g}$    &  $0.128$        &  $0.009$    &        $0.125$  &        $0.005$  &  $0.124$  &   $0.007$  \\
      $\beta_{0,y}+\gamma_y\ln(h_{70})$ & $0.13$ & $0.13$  & $0.09$ & $0.08$ & $0.07$ & $0.12$ \\
      $\beta_{1,y}$ & $1.61$ & $0.04$ & $1.63$ & $0.04$ & $1.63$ & $0.04$ \\
      $\sigma_y$ & $0.182$ & $0.015$ & $0.185$ & $0.016$  & $0.185$ & $0.016$ \\
      \hline
   \end{tabular}
  \end{center}
\end{table}

\begin{figure}
  \centering
  \includegraphics[scale=0.99]{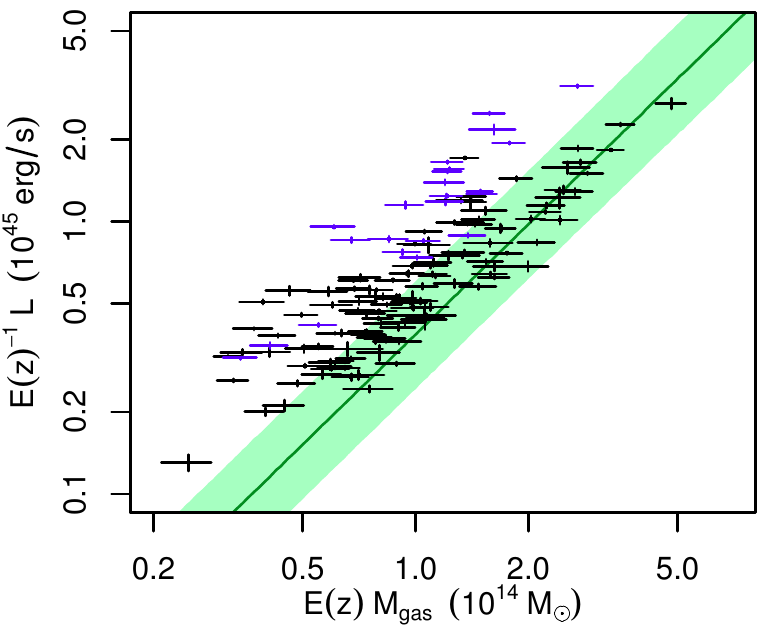}
  \bigskip\\
  \includegraphics[scale=0.99]{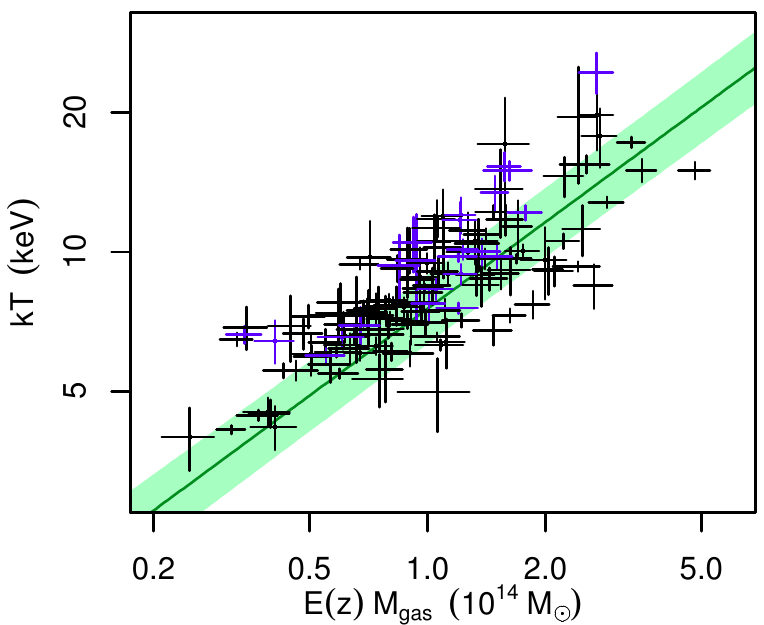}
  \bigskip\\
  \includegraphics[scale=0.99]{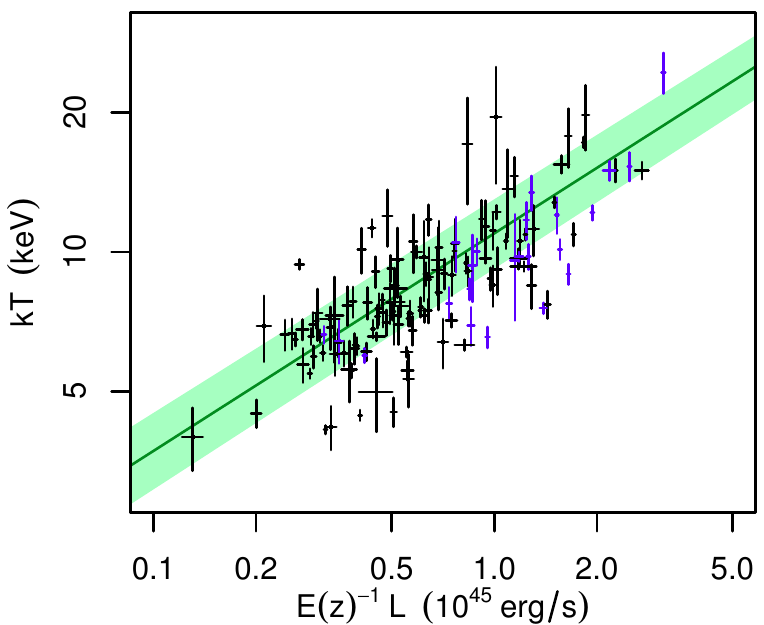}
  \caption{
    Scaling relations of galaxy cluster gas mass, X-ray luminosity and temperature. The values shown are computed within $r_{500}$ for an $h=0.7$, $\Omegam=0.3$ flat \LCDM{} cosmology, and assume $\fgas(r_{500}=0.125)$ (the best fit in this cosmology) in order to determine $r_{500}$. Blue symbols indicate clusters that are classified as relaxed based on their X-ray morphologies. Green lines and shading show the scaling relations fit to the galaxy cluster data for the cosmology above (marginalizing over parameters other than $h$ and $\Omegam$) and their 68.3 per cent predictive intervals, including intrinsic scatter. To translate the scaling relations (which are defined as functions of total mass) into observable space, we consistently assume $\fgas(r_{500})=0.125$.
  }
  \label{fig:xray-scaling}
\end{figure}

\begin{figure*}
  \centering
  \includegraphics[scale=\figscale]{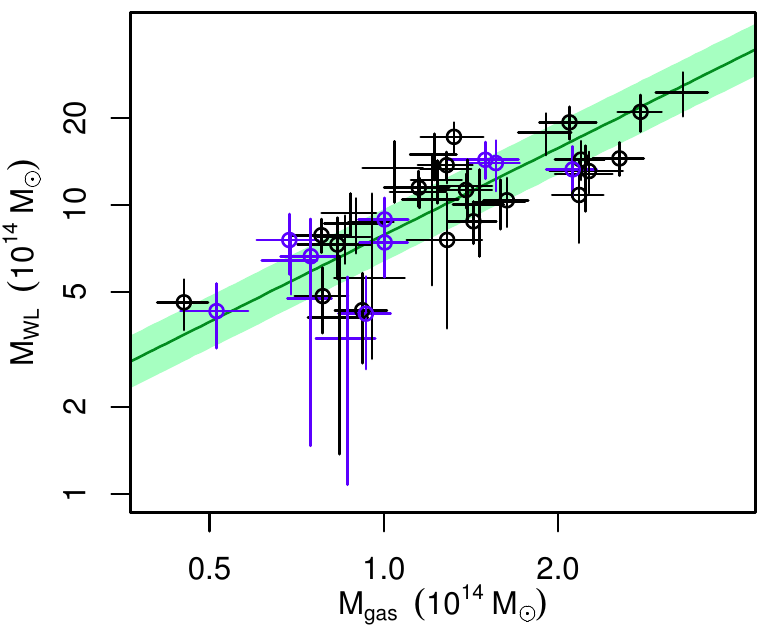}
  \hspace{0.8cm}
  \includegraphics[scale=\figscale]{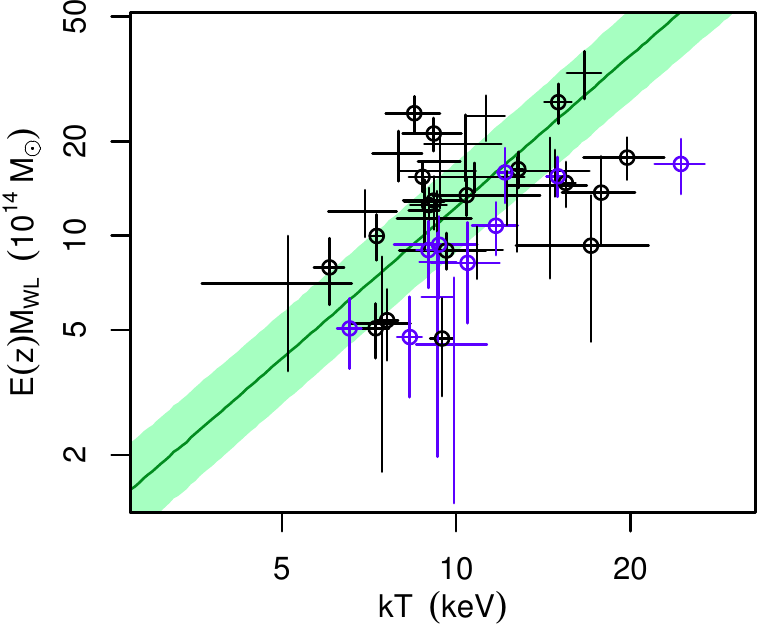}
  \bigskip\\
  \includegraphics[scale=\figscale]{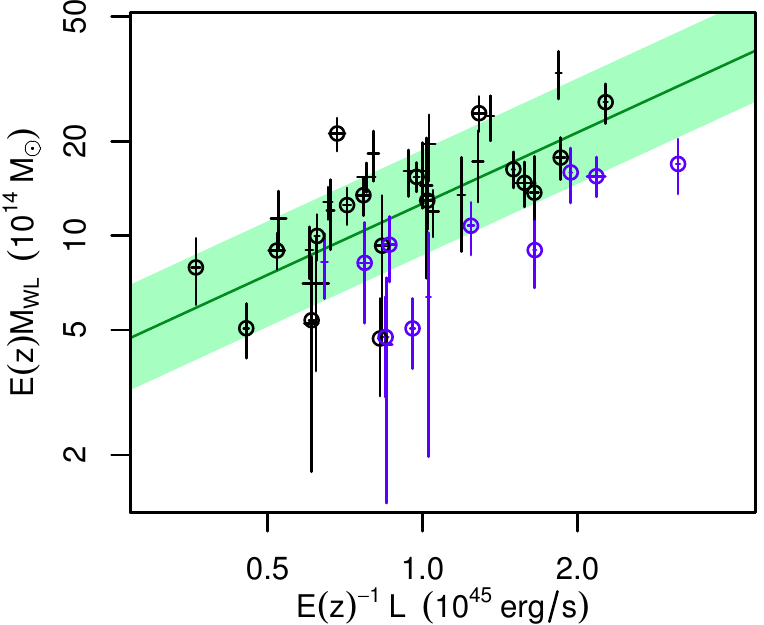}
  \hspace{0.8cm}
  \includegraphics[scale=\figscale]{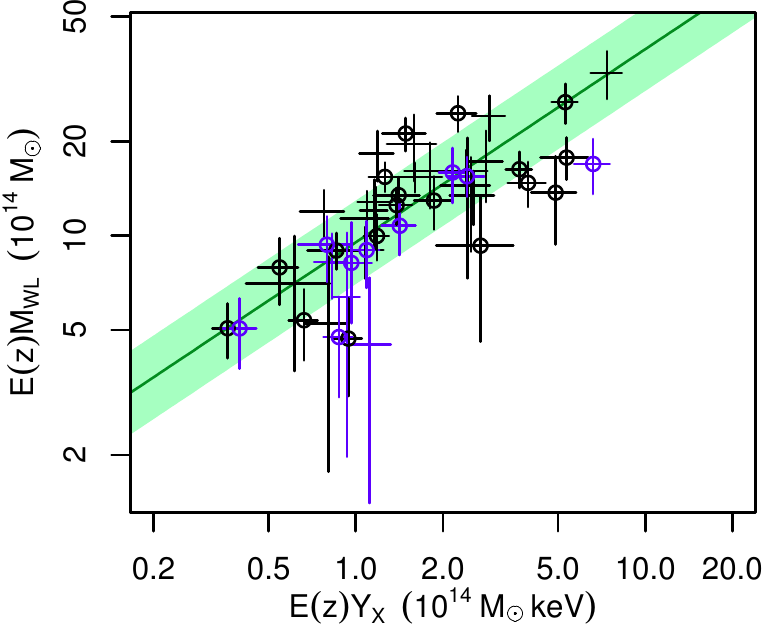}
  \caption{
    Scaling relations of galaxy cluster total mass (from weak lensing), gas mass, X-ray luminosity, temperature, and $\Yx=\Mgas kT$. The values shown are computed within $r_{500}$ for an $h=0.7$, $\Omegam=0.3$ flat \LCDM{} cosmology, and assume $\fgas(r_{500}=0.125)$ (the best fit in this cosmology) in order to determine $r_{500}$. Blue symbols indicate clusters that are classified as relaxed based on their X-ray morphologies. Green lines and shading show the scaling relations fit to the galaxy cluster data for the cosmology above (marginalizing over parameters other than $h$ and $\Omegam$) and their 68.3 per cent predictive intervals, including intrinsic scatter. Here, we plot all clusters with lensing masses determined in Paper~\pthree{}; those that are in the X-ray flux-limited samples used to fit the scaling relations are circled.
  }
  \label{fig:lensing-scaling}
\end{figure*}

Because both the inferred values of luminosity and gas mass, and the inferred scaling relations, are degenerate with the cosmic expansion history, it is most straightforward to compare our measured values with the scaling relation constraints for a restricted cosmological model. Figures~\ref{fig:xray-scaling}--\ref{fig:lensing-scaling} show the measured values derived in our reference $h=0.7$, $\Omegam=0.3$ cosmology with the scaling relations inferred under the same restriction (from Table~\ref{tab:baseline}). For these and the subsequent figures in this section, blue symbols indicate clusters that were identified as hot ($kT>5\keV$) and dynamically relaxed based on their X-ray morphologies by \morph{}, and black symbols show other clusters in our data set. The distinction between relaxed and unrelaxed clusters does not enter into our analysis of the scaling relations in any way, but will be relevant for the discussion in Section~\ref{sec:discussion}. Lines and shading in the figures show the best-fitting scaling relations and 68.3 per cent predictive intervals, accounting for intrinsic scatter. These predictions apply to the underlying cluster population (see Section~\ref{sec:scalingmodel}); that is, we have not re-imposed the selection biases that affect the data onto the model before comparing the two, a feature that is readily apparent in the plot of luminosity and gas mass for this X-ray flux-limited cluster sample (top panel of Figure~\ref{fig:xray-scaling}). The predictions do, however, account for the shape of the mass function, such that, e.g., the predicted range of temperatures at a fixed luminosity includes the range of cluster masses that might give rise to that luminosity, with the proper weighting. Given that only observables are shown here (true mass does not appear), there is an arbitrary choice of which quantity to assign as independent or dependent in every panel, and we have generally followed the convention of treating the quantity with larger measurement uncertainties as dependent for the purposes of these visualizations. The exceptions are the top two panels of Figure~\ref{fig:xray-scaling}, where the small intrinsic scatter of $\Mgas$ allows it to usefully stand in for true mass, providing a relatively direct link to the luminosity and temperature scaling relations in Equation~\ref{eq:ltmgas-scaling}.

The results of our ``clusters only'' and ``all data'' analyses agree within the uncertainties, with the precision being somewhat better in the ``all data'' case, as expected. The constraints are mostly in good agreement with the previous generation of this analysis, \maerd{}, which incorporated $\sim2/3$ as much X-ray follow-up data (with an earlier version of the \Chandra{} calibration) and no weak lensing data. From the ``all data'' analysis, we find a luminosity--mass slope of $1.34\pm0.05$, significantly steeper than the self-similar value of $\sim0.9$ predicted for luminosity in this intrinsic energy range (0.1--2.4\,keV), and a corresponding marginal scatter of $0.43\pm0.03$. Our constraint on the temperature--mass slope is $0.62\pm0.04$, in agreement with the self similar prediction of $2/3$ at the $1\sigma$ level, but markedly different from the \maerd{} result of $0.49\pm0.04$. This shift appears to be the consequence of subsequent changes to the \Chandra{} calibration, specifically as they affect temperature measurements (gas mass measurements agree at the per cent level between the two generations of analysis).\footnote{In addition, we note that \maerd{} used temperatures measured from ASCA data for the lowest-redshift clusters, while we now use \Chandra{} uniformly. While this change may also have an effect, direct comparison of old and new \Chandra{} temperatures indicates that calibration updates primarily account for the change in slope.} The marginal temperature scatter from our analysis is $0.160\pm0.017$. 

The correlation coefficient of the intrinsic scatter in luminosity and temperature at fixed mass is constrained from our analysis to be $0.56\pm0.10$. This positive value is in marginal ($2\sigma$) tension with recent measurements from dynamically relaxed clusters \citep{Maughan1212.0858, Mantz1509.01322} as well as our own analysis using less extensive X-ray follow-up data and an older calibration from Paper~\pfour{}. The astrophysical implications of this measurement will be discussed in Section~\ref{sec:ltcor}.

The temperature--luminosity scaling relation can be obtained by integrating over the mass function, although we obtain essentially identical results by simply algebraically combining the luminosity--mass and temperature--mass relations. For the $h=0.7$, $\Omegam=0.3$ cosmology, the cluster data yield
\begin{equation} \label{eq:tlscaling}
  \expectation{t|\ell} = (1.39\pm0.06) + (0.47\pm0.03)\,\ell,
\end{equation}
with an intrinsic scatter of $0.20\pm0.02$.

Beyond the scaling relations of luminosity and temperature, our analysis also constrains the normalization and slope of the $\Mgas$--$M$ relation. Our constraint on the normalization is equivalent to a gas mass fraction of $\fgas(r_{500})=0.125\pm0.005$ (Table~\ref{tab:baseline}), consistent with our measurements using hydrostatic mass estimates for relaxed clusters \citep{Mantz1509.01322}. The implied gas depletion factor of $\Upsilon_\mathrm{gas}(r<r_{500})=0.80\pm0.08$ (from our $h=0.7$, $\Omegam=0.3$ analysis) is higher than but in reasonable agreement with the values of $\sim0.70$--$0.73$ (independent of redshift) predicted from recent hydrodynamic simulations including radiative cooling, star formation and AGN feedback \citep{Battaglia1209.4082, Planelles1209.5058}. We also obtain a tight constraint on the slope of the $\Mgas$--$M$ relation, $1.007\pm0.012$, consistent with a constant gas-mass fraction.

The temperature and gas mass scaling relations jointly imply a scaling relation for $\Yx=\Mgas kT$, an alternative mass proxy that has been used extensively in recent years (e.g., \citealt{Maughan0703504, Vikhlinin0805.2207, Menanteau1109.0953, Benson1112.5435}). Defining
\begin{equation} \label{eq:yx}
 y_x = \ln\left(\frac{\Mgas{}_{,500} \, kT_{500}}{10^{15}\Msun\keV}\right),
\end{equation}
the nominal scaling relation is
\begin{equation}
  \expectation{y_x} = \left(\beta_{0,\mgas} + \beta_{0,t}\right) + \beta_{1,t}\,\varepsilon + \left(\beta_{1,\mgas} + \beta_{1,t}\right)m.
\end{equation}
Given how closely constrained $\beta_{1,\mgas}$ is to unity, this can be approximated in a more convenient form,
\begin{eqnarray}
  \expectation{y_x}+\varepsilon &\approx& \left(\beta_{0,\mgas} + \beta_{0,t}\right) + \left(\beta_{1,\mgas} + \beta_{1,t}\right)(\varepsilon+m) \nonumber\\
  &\equiv& \beta_{0,y} + \beta_{1,y}(\varepsilon+m).
\end{eqnarray}
The constraints on the normalization, slope and intrinsic scatter parameters for this scaling relation appear in Table~\ref{tab:baseline}, and our $\Yx$ measurements are plotted against weak lensing mass in Figure~\ref{fig:lensing-scaling}.

Finally, we note that, while the normalization and slope of the $\Mlens$--$M$ relation must be determined by priors in our analysis, our data do provide a constraint on the intrinsic scatter of the relation, $\sigma_{\mlens} = 0.17\pm0.06$. This value is consistent with the expected intrinsic scatter in three-dimensional mass estimates from lensing due to triaxiality and projected structure \citep{Becker1011.1681}, and (within large statistical uncertainties) with the scatter between hydrostatic and lensing masses measured from relaxed clusters \citep{Mahdavi1210.3689, Applegate1509.02162}.

\subsection{Comparison with the literature} \label{sec:comparisons}

\begin{figure}
  \centering
  \includegraphics[scale=\figscale]{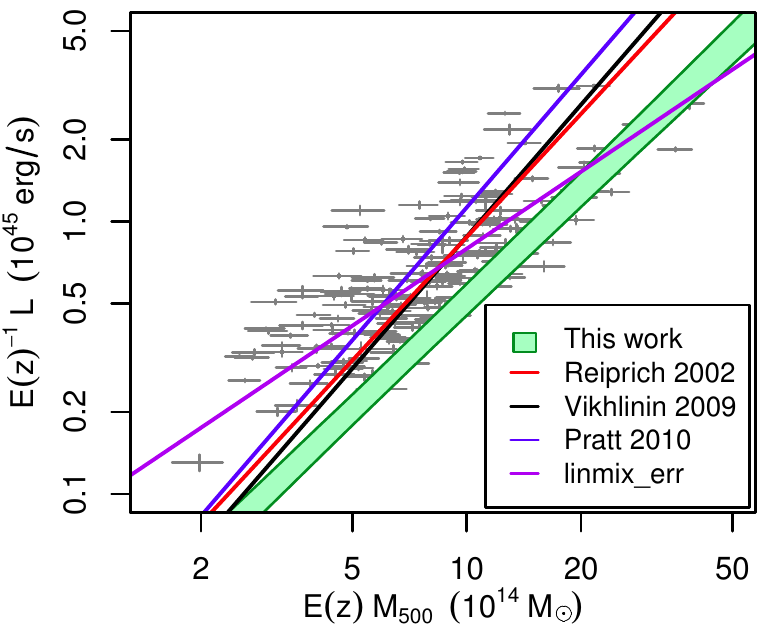}
  \bigskip\\
  \includegraphics[scale=\figscale]{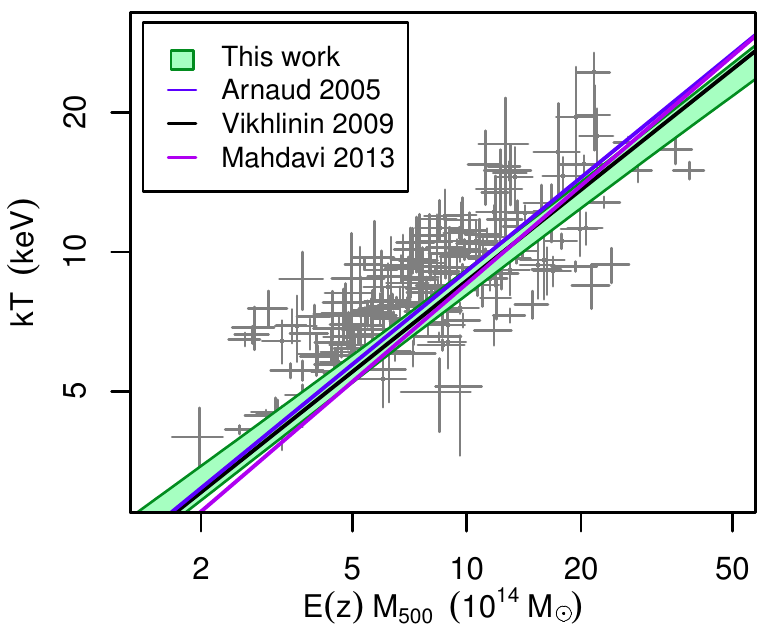}
  \bigskip\\
  \includegraphics[scale=\figscale]{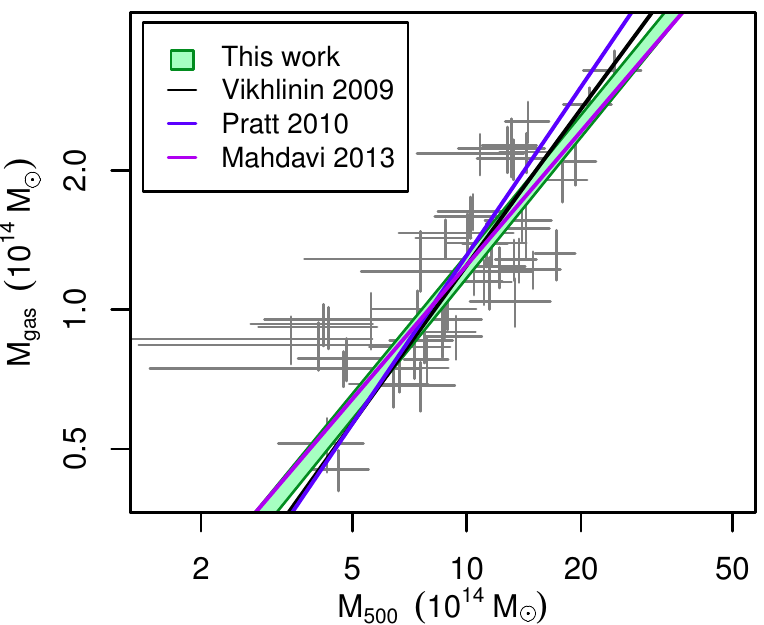}
  \caption{
    Comparison of our fitted scaling relations with others from the literature. Green shading indicates our results for the scaling relations and their 68.3 per cent uncertainty; unlike Figures~\ref{fig:xray-scaling} and \ref{fig:lensing-scaling}, these show only the uncertainty on the nominal scaling relations, and do not include intrinsic scatter. Other lines variously show best fits from \citet{Reiprich0111285}, \citet{Arnaud0502210}, \citet{Vikhlinin0805.2207}, \citet{Pratt0809.3784}, and \citet{Mahdavi1210.3689}. In the top panel, the purple line shows a fit to our data using the regression scheme of \citet{Kelly0705.2774}, which does not account for the selection function or the particular form of the mass function. Since we are comparing scaling relations here, the $M_{500}$ axis label refers to true mass; however, in order to show the range in mass spanned by our follow-up data, we reproduce in the top 2 panels points from Figure~\ref{fig:xray-scaling} (scaling $\Mgas$ to stand in for total mass) and in the bottom panel points from Figure~\ref{fig:lensing-scaling} (showing lensing mass).
  }
  \label{fig:compare}
\end{figure}

In most features, our scaling relations are in good agreement with numerous measurements in the literature (e.g., \citealt{Reiprich0111285, Zhang0702739, Zhang0802.0770, Mantz0709.4294, Mantz0909.3099, Mantz1509.01322, Rykoff0802.1069, Pratt0809.3784, Vikhlinin0805.2207, Leauthaud0910.5219, Reichert1109.3708, Mahdavi1210.3689, Sereno1502.05413}; see also the review of \citealt{Giodini1305.3286}); Figure~\ref{fig:compare} compares our results with some of these. The exception to this good agreement is the luminosity--mass relation, where a clear offset in normalization exists between our results and the others shown (although agreement with \maerd{} is good). We note, however, that the offset is plausibly accounted for by differences in the aperture within which luminosity is measured (up to $\sim5$ per cent), as well as historical differences in mass estimation at the 10 per cent level which are well documented (e.g.\ \citealt{Rozo1204.6301}). The latter affects the luminosity--mass comparison much more strongly than temperature--mass (for hydrostatic mass determinations, mass biases leave the scaling approximately unchanged) or $\Mgas$--mass (due to the weak mass and radius dependence of $\fgas$). The sensitivity of fitted scaling relations to selection biases (\citealt{Vikhlinin0805.2207} and \citealt{Pratt0809.3784} include an accounting for selection effects, although using different methodology than we employ) further complicates this comparison.

The top panel of Figure~\ref{fig:compare} also shows the result of fitting the luminosity--mass relation (with masses estimated from the $\Mgas$ measurements, assuming $\fgas=0.125$) using the {\sc linmix\_err} algorithm of \citet{Kelly0705.2774}, which includes correlated measurement errors and intrinsic scatter, and assumes a log-normal prior on the values of $E(z)M_{500}$. Even though the distribution of masses used in the regression is approximately log-normal (see also discussion in \citealt{Sereno1502.05413}), the clear disagreement between this fit and our primary result demonstrates the danger of adopting a simple prior on cluster masses rather than fully modelling the cosmological mass function and survey selection function. The bias incurred from this approximation is straightforwardly most important for the observable used to select the cluster sample, but at some level affects the scaling of any observable whose intrinsic scatter has non-zero correlation with the selection observable \citep{Allen1103.4829}.

\subsection{Extended models} \label{sec:extended}

Following \maerd{}, we consider parameters that extend the baseline model (see Section~\ref{sec:scalingmodel}), performing a series of analyses in which a single additional parameter is free. For this purpose, we use the external data sets to constrain the cosmological model as much as possible, to maximize leverage on the scaling relation parameters from the cluster data. Table~\ref{tab:extended} shows the results of these tests.

\begin{table}
  \begin{center}
    \caption{
      Best-fitting values and marginalized 68.3 per cent confidence intervals for parameters that extend the baseline scaling relations model. We adopt uniform priors over the ranges shown.
    }
    \label{tab:extended}
    \vspace{1ex}
    \begin{tabular}{cccc}
      \hline
      Parameter & Prior & \multicolumn{2}{c}{Constraint} \\
      & range & 68.3\% & 95.4\% \\
      \hline
      $\beta_{2,\ell}$ & [$-2$, $+2$] & $-0.65\pm0.38$ & $-0.65\pm0.76$ \\
      $\beta_{2,t}$ & [$-2$, $+2$] & $-0.08\pm0.21$ & $-0.08\pm0.44$ \\
      $\sigma'_\ell$ & [$-2$, $+2$] & $-0.61\pm0.24$ & $-0.61\pm0.48$ \\
      $\sigma'_t$ & [$-2$, $+2$] & $1.53_{-0.67}^{+0.47}$ & $1.53_{-1.56}^{+0.47}$ \\
      $\rho'_{\ell t}$ & [$-2$, $+2$] & $-0.5_{-0.7}^{+1.0}$ & $-0.5_{-1.1}^{+2.2}$ \\
      $\lambda_\ell$ & [$-5$, $+5$] & $1.1\pm1.6$ & $1.1\pm3.0$ \\
      \hline
    \end{tabular}
  \end{center}
\end{table}

Compared with \maerd{}, the precision of constraints on these extended models has improved modestly. Whereas the results of that work showed no statistical preferences for any of the model extensions considered, the present analysis constrains two of these parameters, encoding evolution in the marginal intrinsic scatters ($\sigma'_\ell$ and $\sigma'_t$), to be non-zero at the $\sim2\sigma$ level. The latter should be treated with some caution, since the data provide only a lower bound on $\sigma'_t$, with the posterior distribution becoming flat for positive values. Nevertheless, it is interesting that the tentative evidence we have for departures from the baseline model favors evolution in the marginal intrinsic scatters rather than in their correlation or in the normalizations of the scaling relations. We discuss the astrophysical implications of these results in Section~\ref{sec:scatevol}.

Several authors have considered departures from self-similar evolution in the scaling relations similar to that encoded by the parameters $\beta_{2,\ell}$ and $\beta_{2,t}$ here. \citet{Ettori0312239} and \citet{Morandi0704.2678} find no evidence for non-similar evolution in either the luminosity or temperature scaling relations; their data sets respectively span $0.4<z<1.3$ and $0.14<z<0.82$. \citet[][$0.0<z<0.9$]{Vikhlinin0805.2207} and \citet[][$0.0<z<1.1$]{Mantz1509.01322} find marginally weaker evolution in luminosities than the self-similar model, qualitatively in line with our results, while \cite[][$0.0<z<1.5$]{Reichert1109.3708} and \citet[][$0.0<z<1.5$]{Sereno1502.05413} find substantially more negative evolution in luminosities. Note that the apparent evolution of the luminosity--mass relation is highly sensitive to X-ray flux selection; some of the variation in these results is likely to be due to selection biases.

\section{From Scaling Relations to Astrophysics} \label{sec:discussion}

In this section, we discuss the astrophysical consequences of the scaling relation constraints presented in Section~\ref{sec:results}. Section~\ref{sec:slopes} addresses the slopes of the scaling relations, while Sections~\ref{sec:ltcor} and \ref{sec:scatevol} concern the intrinsic covariance of luminosity and temperature at fixed mass, its origin, and its evolution.

\subsection{Scaling relation slopes} \label{sec:slopes}

The slopes of the temperature and gas mass scaling relations from our analysis are in good agreement with the self-similar model of cluster formation through spherical collapse \citep{Kaiser1986MNRAS.222..323K}, which predicts values of $2/3$ and unity,  respectively. The slope of the luminosity--mass relation, in contrast, is steeper than self-similar; accounting for the limited energy band used in our definition of $L$ (0.1--2.4\,keV in the rest frame), the self-similar prediction corresponds to a slope of $\sim0.9$ for hot clusters. This departure is well documented \citep{Giodini1305.3286}, and previous studies showed that self-similar behavior is recovered if a central region of radius $\sim0.15\,r_{500}$ is excised from the luminosity measurement (\citealt{Maughan0703504, Zhang0702739}; \maerd{}; \citealt{Mantz1509.01322}). In this context, recall that the temperature measurements used here, which result in a nearly self-similar scaling relation, are also center-excised. Gas mass, like luminosity, is essentially an integral of the observed surface brightness, but, being weighted by radius, it is naturally less sensitive to cluster centers. Consequently and consistently, the scaling relation slope of gas mass within $r_{500}$ is self-similar, as is $\Mgas$ measured in a spherical shell (excluding the center) at smaller radii, while the gas mass integrated within $\sim0.2\,r_{500}$ has a clearly non-similar slope \citep{Mantz1509.01322}.

The data thus indicate that, outside of cluster centers, the ICM in massive clusters is well described by the simple self-similar model. Self-similarity is only strongly broken, presumably by some combination of feedback and radiative cooling, in a limited volume of clusters near their centers. In recent work \citep{Mantz1509.01322} we reached much the same conclusions based on an analysis of exclusively dynamically relaxed clusters; here we show that the same statements (and the same slopes) hold for a larger, dynamically heterogenous cluster sample, and when fully accounting for selection biases.

\subsection{Correlation and origin of $L$ and $kT$ scatter} \label{sec:ltcor}

\begin{figure*}
  \centering
  \includegraphics[scale=\figscale]{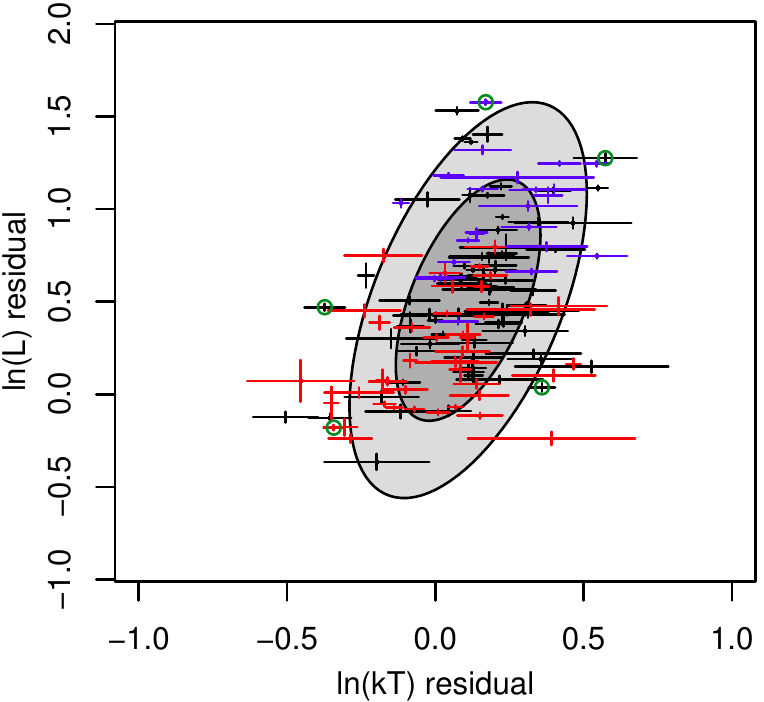}
  \hspace{1mm}
  \includegraphics[scale=1]{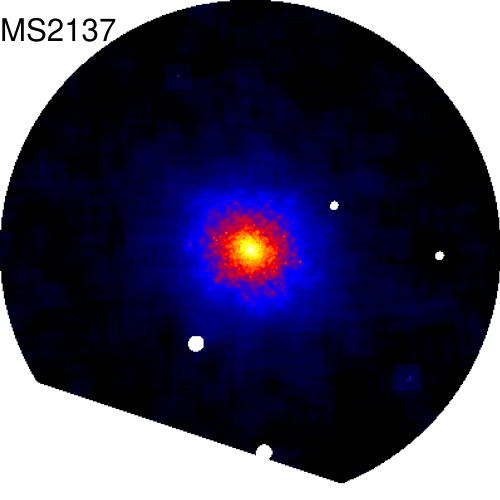}
  \hspace{1mm}
  \includegraphics[scale=1]{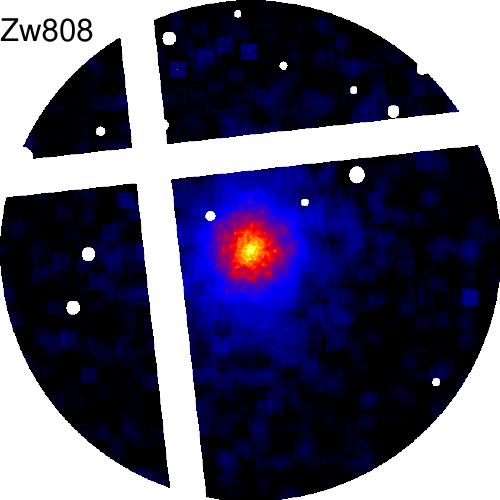}\vspace{10mm}\\
  \includegraphics[scale=1]{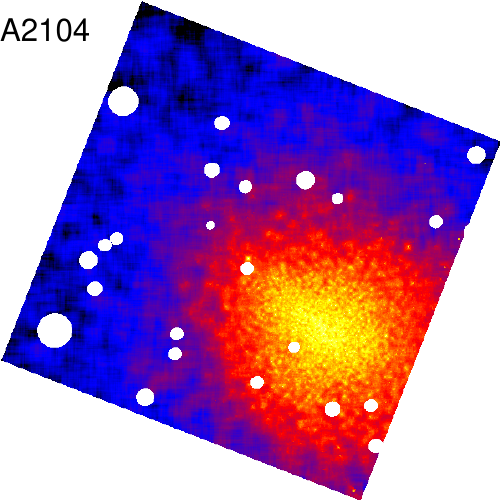}
  \hspace{5mm}
  \includegraphics[scale=1]{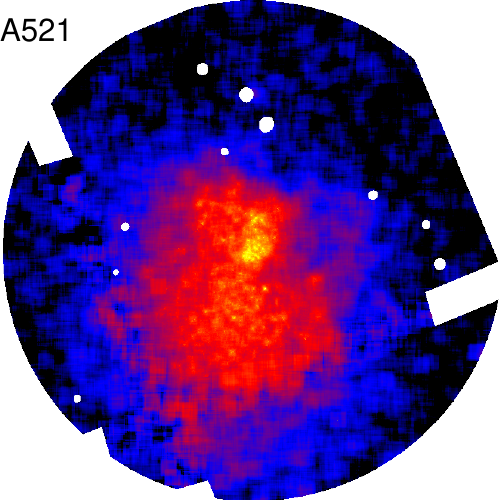}  
  \hspace{5mm}
  \includegraphics[scale=1]{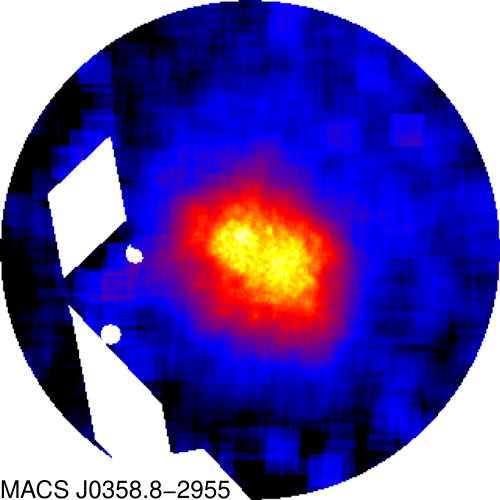}
  \caption{
    Top-left: distribution of luminosity and temperature residuals from the best-fitting scaling relation, from the clusters-only analysis with fixed $h$ and \Omegam{}. Here, cluster gas masses are used to determine the nominal total mass, and hence the residuals. Blue points correspond to highly relaxed clusters, red clusters are classified as likely mergers based on their X-ray morphology or the presence of a radio halo or relic, and black points fall into neither of these categories. Gray, shaded ellipses show the 68.3 and 95.4 per cent confidence regions associated with the correlated intrinsic scatter in luminosity and temperature at fixed mass; the centers of these have been offset from the origin purely to emphasize the similarity between the shapes of these ellipses and the distribution of measured residuals. X-ray images of representative clusters from the periphery (circled in green) are shown in the remaining panels; clockwise from the top of the residual plot, these are (1) MS\,2137, (2) Zwicky~808, (3) Abell~2104, (4) Abell~521, and (5) MACS\,J0358.8$-$2955.
  }
  \label{fig:LTresid}
\end{figure*}

Our baseline analysis finds a positive correlation in the intrinsic scatter of luminosity and temperature at fixed mass, with $\rho_{\ell t} = 0.56\pm0.10$, a result consistent with predictions from hydrodynamical simulations \citep{Stanek0910.1599}, but which is marginally at odds with constraint of $-0.06\pm0.24$ for relaxed clusters from \citet{Mantz1509.01322}.\footnote{\citet{Maughan1212.0858} found a correlation of $0.37\pm0.31$ between temperature and \emph{bolometric} luminosity for a dynamically heterogenous sample. Assuming $\sigma_\ell=0.43$ and $\sigma_t=0.16$, the $T^{1/2}$ dependence of the bolometric luminosity should naturally impose a correlation coefficient of $\sim0.2$ in the scatter. The \citet{Maughan1212.0858} result is thus more consistent with zero correlation of the temperature and soft-band luminosity than with our nominal value of 0.54, although the uncertainty is large enough that the two are stil formally consistent. In contrast, the \citet{Stanek0910.1599} correlation estimate of $\sim0.7$ between bolometric luminosity and spectroscopic temperature is compatible with our constraint once adjusted to the soft band.} Insight into this discrepancy comes from the distribution of residuals in $L$ and $kT$ from their best-fitting scaling relations. The first panel of Figure~\ref{fig:LTresid} shows these residuals, where $\Mgas$ has been used as a mass proxy to determine the nominal values of $L$ and $kT$ for each cluster, and where relaxed clusters are shown in blue.\footnote{Recall that this sub-sample is defined using a temperature cut ($kT>5$\,keV) in addition to a morphological test. While a strictly morphological selection would be more inclusive and slightly more relevant for this discussion, we retain the same definition for consistency with previous sections, and because this is the selection used in the scaling relation analysis of \citet{Mantz1509.01322}, which we compare to.} Confidence regions at the 68.3 and 95.4 per cent levels associated with the best-fitting covariance are overlaid. Visually, there is indeed a positive trend of the residuals, consistent with the fitted model, when considering all of the clusters. The relaxed sub-sample, however, occupies a smaller area of the plot, being confined to higher than average luminosities and approximately average temperatures. In particular, the tail extending to low luminosities and temperatures is excluded, reducing the $L$--$kT$ correlation of the subsample. These relatively lower luminosity and lower temperature systems are, in contrast, preferentially mergers; clusters shown as red points in the figure either host known radio halos or relics \citep{Feretti1205.1919, Cassano1306.4379} or have disturbed X-ray morphologies. For this purpose, ``disturbed'' means that a cluster would be classified as unrelaxed based on both the ``symmetry'' and ``alignment'' metrics introduced by \morph{}. Images of outlying clusters, shown in the remaining panels of Figure~\ref{fig:LTresid}, reinforce this impression.

\begin{figure*}
  \centering
  \includegraphics[scale=\figscale]{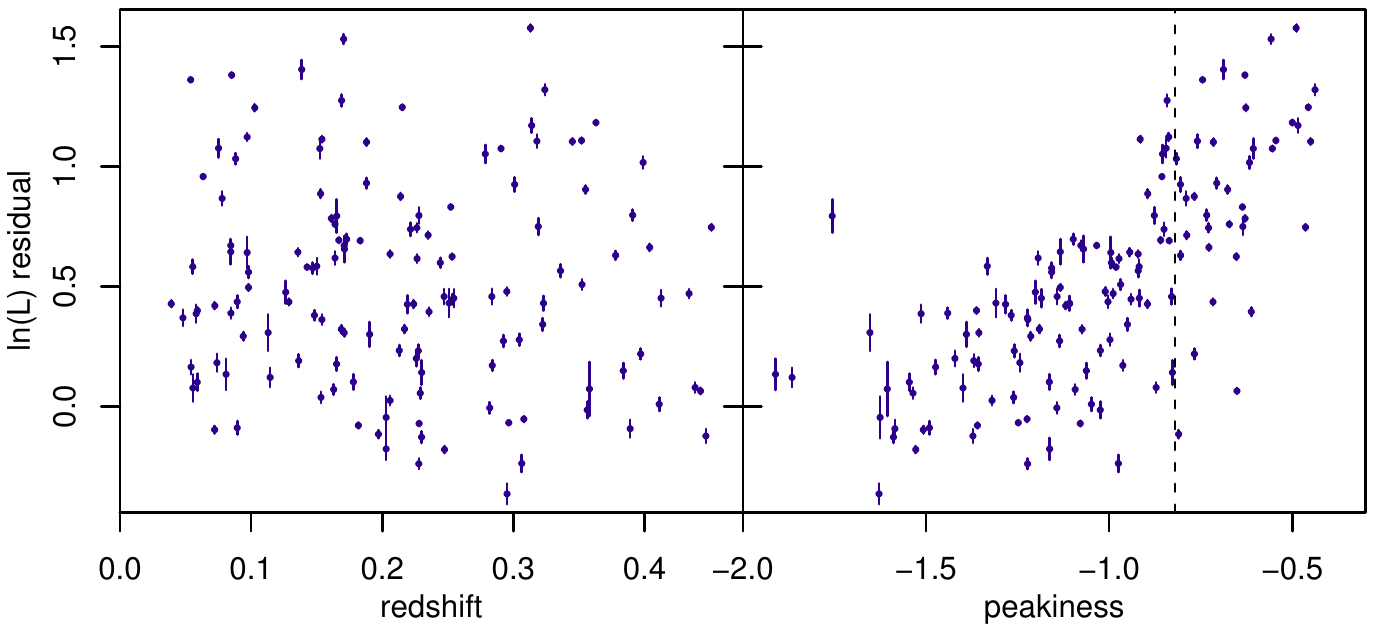}\vspace{5mm}\\
  \includegraphics[scale=\figscale]{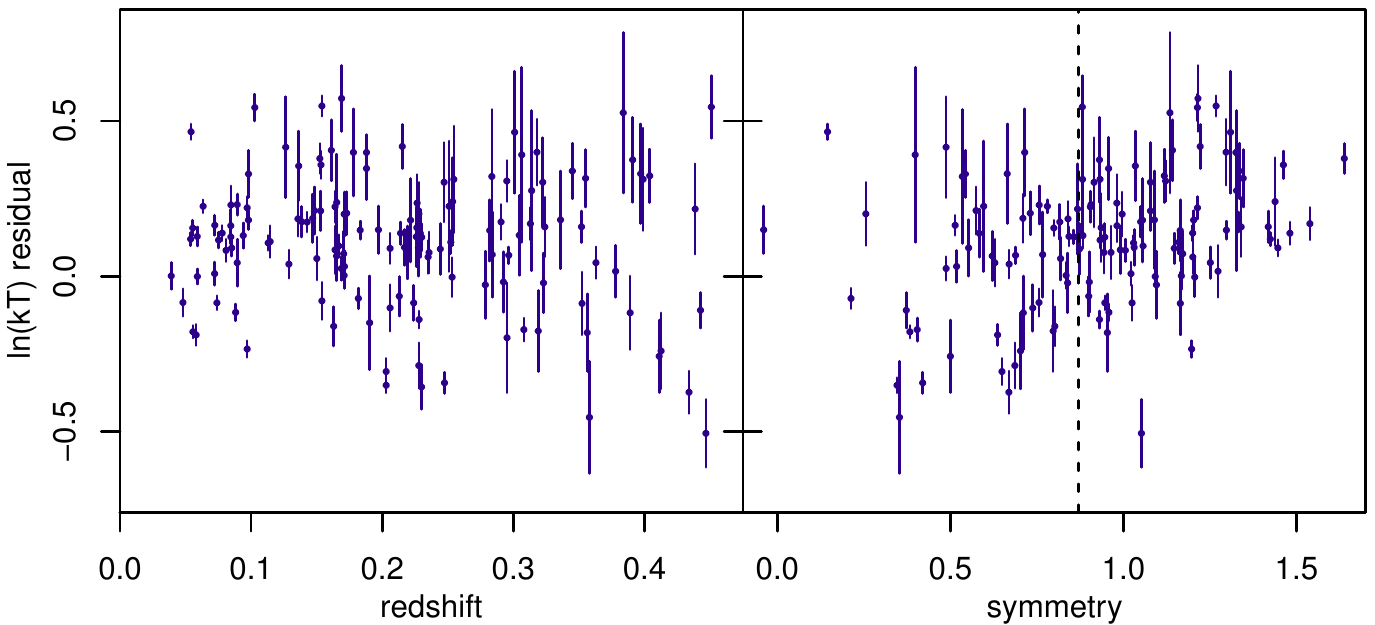}
  \caption{
    Luminosity and temperature residuals, as in Figure~\ref{fig:LTresid}, are plotted against redshift and either the ``peakiness'' or ``symmetry'' morphological metrics of \morph{}. These morphological indicators are, respectively, related to the sharpness of the surface brightness peak (i.e.\ cool, bright cores) and the alignment of the brightness peak with lower brightness isophotes (i.e.\ an overall undisturbed appearance).
  }
  \label{fig:scatzmorph}
\end{figure*}

The partial segregation of relaxed and merging clusters in this plane has a simple physical interpretation. During a merger, the total mass and gas mass of the composite cluster increase essentially immediately. The temperature does not, however, apart from the possibility of a transient shock. The global cluster temperature will not approach the final value appropriate to its new mass until residual bulk motions of the gas have been virialized \citep{Mathiesen0004309, Kravtsov0603205, Ventimiglia0806.0850, Rasia1012.4027}. The luminosity, similarly, takes some time to approach its new equilibrium value, even though the global luminosity is boosted proportionately by the addition of the subcluster during the merger. We therefore might expect clusters undergoing major mergers to preferentially be scattered low in luminosity and temperature, given their total masses, compared to the cluster population at large. We might also expect merging clusters to have a larger temperature scatter overall, given the possible presence of shocks. In contrast, the temperatures of dynamically relaxed clusters should most tightly track their total masses, since they are closest to virial equilibrium. Given that the presence of a bright, cool core is one of the criteria for relaxation employed here, we also expect the relaxed sample to occupy higher-than-average luminosities.

We can see these individual trends at work in Figure~\ref{fig:scatzmorph}, which plots the luminosity and temperature residuals against the ``peakiness'' and ``symmetry'' morphological metrics of \morph{}, respectively. In the case of the luminosity residuals, there is a clear trend with peakiness, which measures how centrally concentrated a cluster's surface brightness is, consistent with previous indications that the luminosity scatter is driven by cluster centers (\citealt{Maughan0703504, Zhang0702739}; \maerd{}; \citealt{Mantz1509.01322}). The peakiness and symmetry thresholds required for a cluster to be classified as relaxed are shown by dashed lines. Although it is less strong, there is also a net positive trend relating symmetry and the temperature residuals, consistent with less relaxed clusters being cooler at a given mass.

The dynamical-state dependence of scaling relation residuals is also clear in the individual plots in Figures~\ref{fig:xray-scaling} and \ref{fig:lensing-scaling}. In particular, in Figure~\ref{fig:lensing-scaling}, the relaxed clusters consistently have high values of temperature, given their lensing masses. The same dependence on dynamical state is not apparent in the residuals from the lensing mass--gas mass relation. The plot of lensing mass against $\Yx$ is an intermediate case, but still displays much of the bias evident in the mass--temperature panel. This observation suggests that the anti-correlation in $\Mgas$ and $kT$ residuals expected from simulations, which would reduce the offset between relaxed and unrelaxed clusters \citep{Kravtsov0603205}, may not hold for very massive clusters. If true, this dynamical-state bias would pose a problem for the use of $\Yx$ as a mass proxy when calibrated using hydrostatic masses for relaxed clusters (e.g.\ discussion in \citealt{Vikhlinin0805.2207}).

\subsection{Evolution of scatter} \label{sec:scatevol}

In Section~\ref{sec:extended}, we found tentative evidence for evolution in the intrinsic scatters of both luminosity and temperature at fixed mass.\footnote{Note that, since gas-mass fraction measurements are consistent with a constant intrinsic scatter \citep{Mantz1402.6212}, this also tentatively points to an increasing scatter in $\Yx$ with redshift.} In an analysis with both of these evolution parameters free, they are weakly correlated a posteriori, indicating that the evidence for evolution in each quantity is due to independent features of the data. The individual luminosity and temperature residuals from the best-fitting model are shown against redshift in Figure~\ref{fig:scatzmorph}.

Astrophysically, the preference for evolving scatter can be understood in the same terms as the scatter correlation in Section~\ref{sec:ltcor}. Namely, if the marginal scatter in temperature at fixed mass is driven by mergers, then an increasing scatter with redshift points to a larger merger rate for massive clusters at higher redshifts. Interestingly, however, no clear redshift dependence is seen in the fraction of morphologically disturbed clusters, whether selected by X-ray emission or the Sunyaev-Zeldovich (SZ) effect (\morph{}; though see also \citealt{Mann1111.2396}). A more complete understanding of how temperature and morphology-based diagnostics respond to cluster dynamics is clearly desirable. In terms of the luminosity, the preference for decreasing scatter presumably reflects the late-time development of dense, bright cores in a subset of the population (e.g.\ \citealt{Santos0802.1445, Santos1008.0754, McDonald1305.2915}; \morph{}). This interpretation is also consistent with the preference, albeit statistically weaker, for negative evolution in the mean luminosity--mass relation in Section~\ref{sec:extended}.

Using hydrodynamical simulations, \citet{Stanek0910.1599} and \citet{Le-Brun1606.04545} find approximately constant intrinsic scatters in luminosity and temperature, at high masses and redshifts $<0.5$. In the case of luminosity, where our evidence for evolution is stronger, this may reflect the challenge of reproducing realistic cool cores in simulations.

\section{Scaling of Non-X-ray Proxies} \label{sec:szo}

This section presents scaling relations of our X-ray and lensing mass proxies with measurements of optical richness ($\lambda$) from the Sloan Digital Sky Survey (SDSS) redMaPPer catalog (version 5.10;\footnote{\url{http://risa.stanford.edu/redmapper/}} \citealt{Rykoff1303.3562, Rozo1410.1193}) and of Compton $Y$ from the \citet{Planck1502.01598}. We consider these results separately from those of Section~\ref{sec:results} because the observables $\lambda$ and $Y$ are not presently incorporated into the simultaneous cosmology+scaling relation fitting code used elsewhere in this work; instead, we fit the scaling relations employing simpler Bayesian regression methods \citep{Kelly0705.2774, Mantz1509.00908}, and using quantities derived for a fixed flat \LCDM{} ($h=0.7$, $\Omegam=0.3$) cosmology. Consequently, selection effects induced by our X-ray flux-limited sample, or the availability of external $\lambda$ and $Y$ measurements, are not accounted for. While we expect the selection effects to have a minor impact on scaling relations that do not involve X-ray luminosity, our results in this section should be interpreted with this in mind.

\subsection{redMaPPer richness}

\begin{figure}
  \centering
  \includegraphics[scale=\figscale]{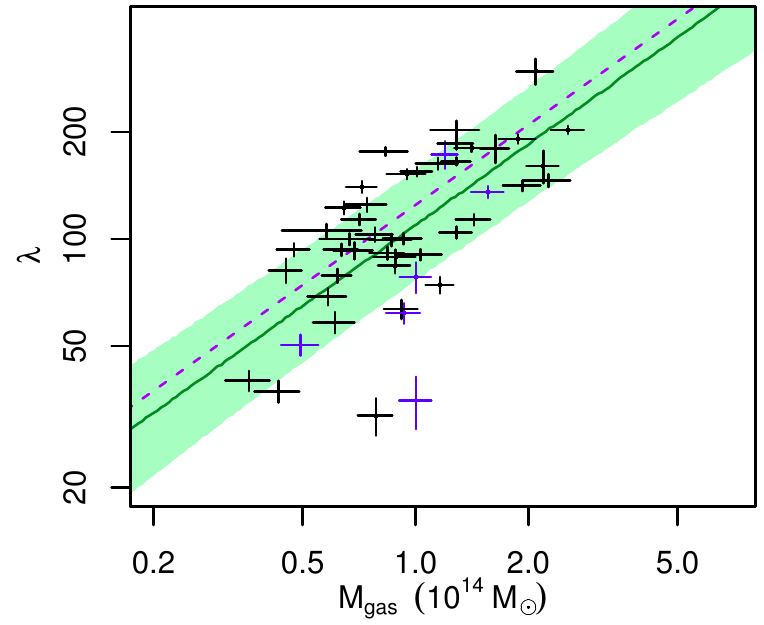}
  \\
  \includegraphics[scale=\figscale]{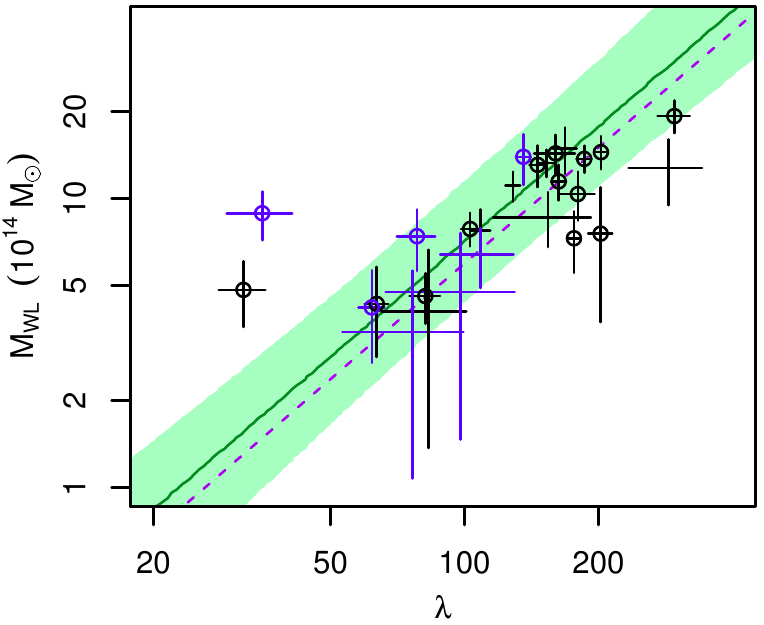}
  \\
  \includegraphics[scale=\figscale]{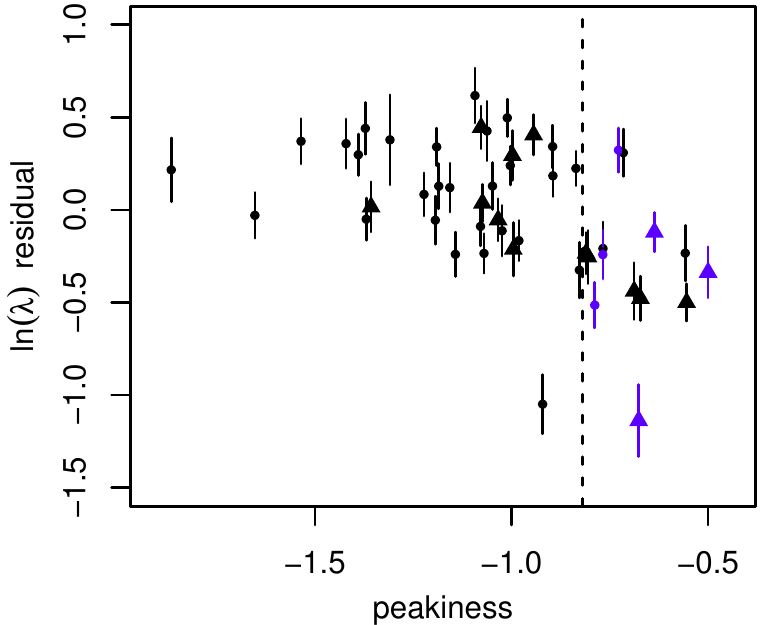}
  \caption{
    Top and center: scaling relations of our X-ray and weak lensing cluster measurements with redMaPPer optical richness \citep{Rykoff1303.3562, Rozo1410.1193}. Points and green lines/shading are as in Figures~\ref{fig:xray-scaling} and \ref{fig:lensing-scaling}. The purple, dashed lines show the $\lambda$--mass relation determined by \citet{Simet1603.06953} using stacked weak lensing of SDSS redMaPPer clusters. 
    Our fits as shown use all of the clusters plotted; when mis-centered and otherwise problematic clusters are excluded (see below) the normalization of our scaling relation is in agreement with that of \citet{Simet1603.06953} at the few per cent level.
    Bottom: log-residuals from the mean $\lambda$--$\Mgas$ relation are plotted against X-ray surface brightness peakiness from \morph{}. The vertical, dashed line corresponds to the threshold peakiness for a cluster to be considered relaxed in that work. Triangles indicate clusters where redMaPPer chooses an incorrect CG (see text). The remaining large outlier is Abell~963, a known case where photometric errors significantly bias the richness measurement.
  }
  \label{fig:rmp}
\end{figure}

In Figure~\ref{fig:rmp}, we plot redMaPPer richness against our gas mass and lensing mass measurements. Fitting the former scaling relation, we find
\begin{equation}
  \lambda = e^{4.69\pm 0.05} \left[\frac{\Mgas{}_{,500}}{10^{14}\Msun}\right]^{0.75\pm 0.12},
\end{equation}
with a log-normal intrinsic scatter of $0.35\pm 0.04$. Using our $\Mgas$--$M$ relation to interpret this in terms of total mass, the constraints on the slope and normalization are in excellent agreement with recent results from \citet[][dashed line in the figure\footnote{The \citet{Simet1603.06953} scaling relation is given in terms of an overdensity of 200 with respect to the mean matter density of the Universe, as opposed to our convention of 500 with respect to the critical density. We have re-normalized their results using a conversion factor based on the \citet*{Navarro9611107} mass profile, with an appropriate concentration parameter for the SDSS redMaPPer sample, as given by \citet{Rykoff1104.2089}.}]{Simet1603.06953}, who used stacked weak lensing for SDSS clusters with $20\leq\lambda\leq140$ to constrain the $\lambda$--$M$ relation. We note, however, that the intrinsic scatter in richness from our analysis is somewhat greater than the initial estimates of 0.2--0.3 by \citet{Rykoff1303.3562}, which were incorporated into the \citet{Simet1603.06953} analysis as a prior (but see below).

While there are only 6 relaxed clusters in the comparison sample in the top panel of Figure~\ref{fig:rmp}, we can tentatively identify a tendency for relaxed clusters to have relatively low $\lambda$ for their masses. As the relaxed subsample contains undisturbed clusters with strong cool cores (by construction) and consequently relatively high central-galaxy (CG) star formation rates (e.g.\ \citealt{Rafferty0802.1864}), the difficulty of correctly identifying star-forming CG's using red-sequence methods plausibly contributes to this trend. The bottom panel of Figure~\ref{fig:rmp} reinforces this impression, showing a trend in the $\ln(\lambda)$ residuals from the best-fitting relation with the sharpness of the X-ray surface brightness peak (from \morph{}). Triangles in the figure identify clusters where the redMaPPer-assigned CG differs from the CG chosen by eye by \morph{} by $>50$\,kpc. While mis-centering at this level occurs throughout the sample, it is most prevalent among the X-ray peakiest clusters; the rate of mis-centering is almost $3\times$ greater for clusters with peakiness $>-0.82$ than for less peaky clusters. The clear outlier that is not mis-centered is Abell~963, a case where the redMaPPer richness is known to be strongly affected by photometric errors (E.\ Rozo, private communication).

If we discount the mis-centered clusters and Abell~963, there is little evidence for any remaining trend in the $\lambda$--$\Mgas$ residuals with X-ray peakiness. Fitting the scaling relation to this more restricted data set, we find
\begin{equation}
  \lambda = e^{4.80\pm 0.05} \left[\frac{\Mgas{}_{,500}}{10^{14}\Msun}\right]^{0.73\pm 0.11},
\end{equation}
with a log-normal scatter of $0.27\pm0.04$. In addition to having a smaller scatter than the previous scaling relation, the normalization of this new fit is in excellent agreement with that of \citet{Simet1603.06953}; at $M=10^{15}\Msun$, the mean $\lambda$ values predicted by the two relations differ by 3 per cent. We note that \citet{Farahi1601.05773} also find a scaling relation compatible with that of \citet{Simet1603.06953} using stacked velocity dispersions to estimate mass.

\subsection{\Planck{} Compton $Y$}

\begin{figure}
  \centering
  \includegraphics[scale=\figscale]{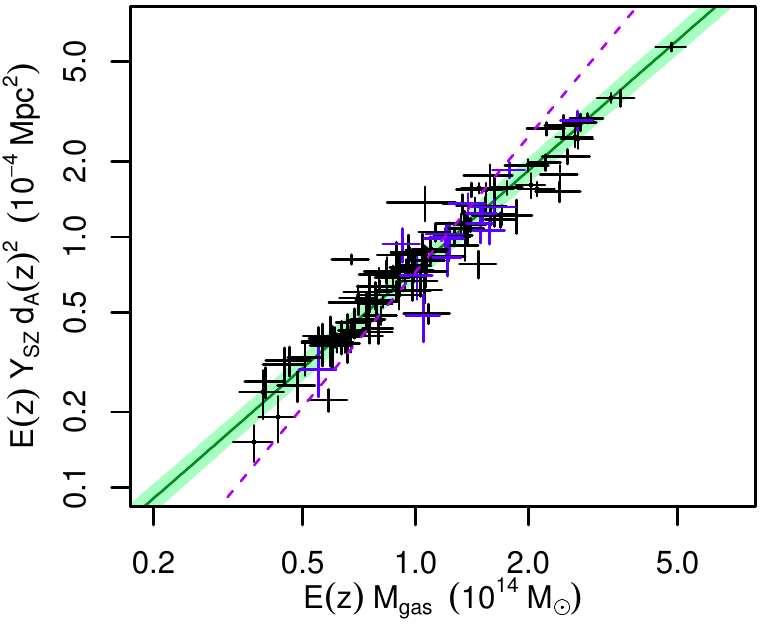}
  \bigskip\\
  \includegraphics[scale=\figscale]{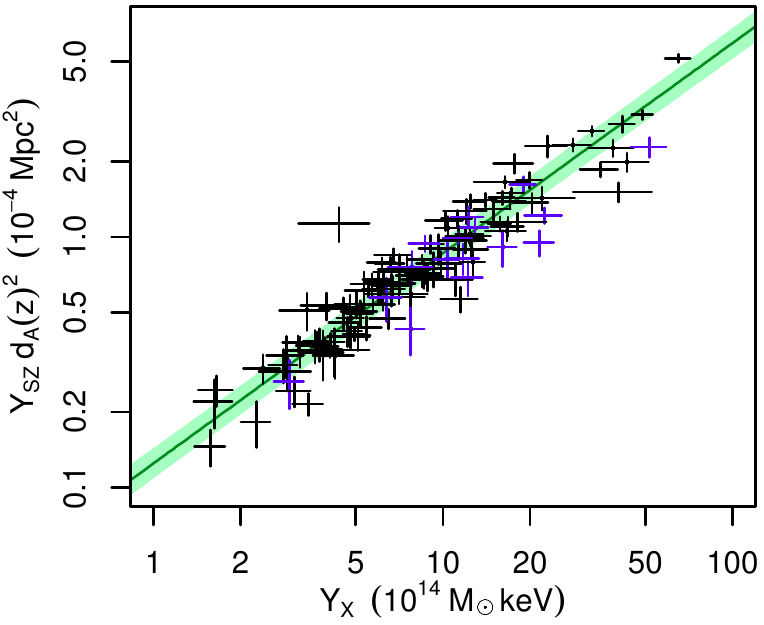}
  \bigskip\\
  \includegraphics[scale=\figscale]{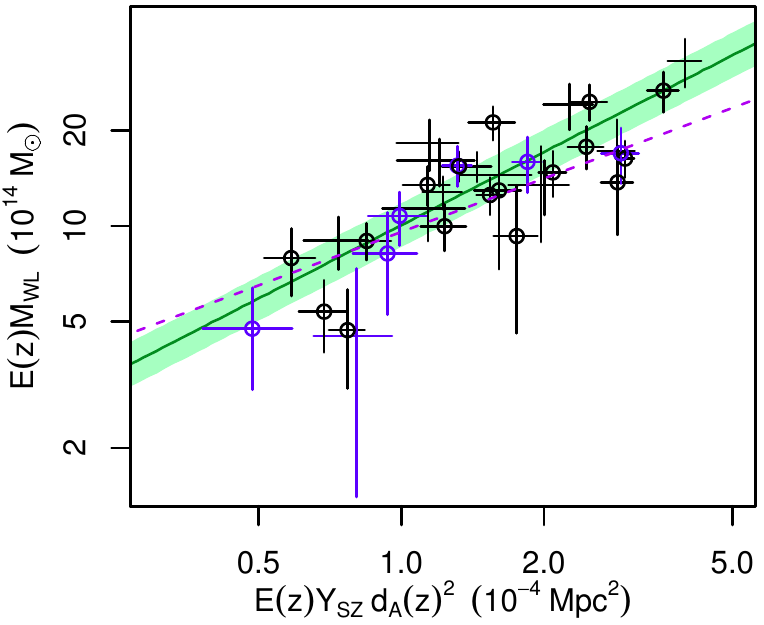}
  \caption{
    Scaling relations of our X-ray and weak lensing cluster measurements with spherically integrated Compton $Y$ from the \citet{Planck1502.01598}. Points and green lines/shading are as in Figures~\ref{fig:xray-scaling} and \ref{fig:lensing-scaling}. The purple, dashed lines show the $Y$--mass relation assumed in the Planck cluster cosmological analysis \citep{Planck1303.5080, Planck1502.01597}. Note that the green shading in the bottom panel is based on a fit to the $Y$--$\Mgas$ data in conjunction with the $\Mgas$--$M$ relation, not a direct fit to the $Y$--$\Mlens$ data.
  }
  \label{fig:sz}
\end{figure}

The \citet{Planck1502.01598} fits a generalized NFW pressure model to each cluster they detect via the SZ effect, varying only the scale radius and spherically integrated Compton $Y$ within $5\,r_{500}$. Given the fixed shape of the pressure model, these quantities are deterministically related to $r_{500}$ and $Y(r_{500})$, respectively. We importance sample the Planck fits (specifically from the MMF3 catalog\footnote{\url{http://irsa.ipac.caltech.edu/data/Planck/release_2/catalogs/}}), using our constraints on $r_{500}$, to arrive at a constraint on $Y_{500}$ for each cluster in common. These values are plotted against our $\Mgas$, $\Mlens$ and $\Yx$ measurements in Figure~\ref{fig:sz}.

Fitting the Planck Compton $Y$ measurements against gas mass, we obtain the scaling relation
\begin{equation} \label{eq:Ysz-M}
  \frac{E(z)\,Y_{500}\,\dA(z)^2}{10^{-4}\,\Mpc^2} = e^{-0.299\pm 0.018} \left[\frac{E(z)\,\Mgas{}_{,500}}{10^{14}\Msun}\right]^{1.31\pm 0.03},
\end{equation}
with a log-normal intrinsic scatter of $0.117\pm 0.018$,\footnote{Note that the intrinsic covariance of $Y$ and $\Mgas$ at fixed mass is expected to be positive (e.g.\ \citealt{Stanek0910.1599}), in which case the marginal intrinsic scatter in $Y$ will be larger than this value.} where $\dA(z)$ is the angular diameter distance. Considered in light of our nearly linear $\Mgas$--$M$ relation, this slope is in significant tension with the scaling of $Y\,\dA^2 \propto M_\mathrm{Pl}^{1.79}$ assumed by the \citet{Planck1303.5080, Planck1502.01597} in their cosmological analysis. Evidence of a mass-dependent offset between weak lensing masses and masses estimated from Planck $Y$ measurements and their assumed scaling relation has previously been noted \citep{von-der-Linden1402.2670, Hoekstra1502.01883}. Our results here are consistent, implying $M_\mathrm{Pl} \propto M^{0.730\pm0.019}$, compared with $M_\mathrm{Pl} \propto M^{0.68^{+0.15}_{-0.11}}$ from the direct comparison of WtG and Planck masses \citep{von-der-Linden1402.2670}.\footnote{However, we note one difference in these two results: in this work we extract masses and Compton $Y$ measurements using a consistent value of $r_{500}$, whereas the comparison of weak lensing and \Planck-derived masses allowed each measurement to correspond to a different radius.} In addition to improving the precision of this measurement, the larger comparison sample employed here, compared with that of \citet{von-der-Linden1402.2670}, allows us to disentangle trends with mass from potentially competing trends with redshift. Fitting the data to a power law of the form $Y_{500}\,\dA(z)^2 \propto \Mgas^{\alpha_1}\,E(z)^{\alpha_2}$, we find no preference for evolution beyond that present in Equation~\ref{eq:Ysz-M}, with $\alpha_2-(\alpha_1-1) = 0.02\pm0.39$. Interestingly, and unlike the case of $\Yx$ (Section~\ref{sec:baseline}), the relaxed subsample of clusters does not preferentially scatter in one direction from the mean $Y_{500}$--mass relation.

The difference between our derived $Y$--$M$ and $\Yx$--$M$ slopes implies a $Y$--$\Yx$ relation that strongly departs from linearity. This is verified in Figure~\ref{fig:sz}; we obtain
\begin{equation} \label{eq:y-yx}
  \frac{Y_{500}\,\dA(z)^2}{10^{-4}\,\Mpc^2} = e^{-0.338\pm 0.019} \left[\frac{\Yx}{8\E{14}\Msun\keV}\right]^{0.84\pm 0.03},
\end{equation}
with a log-normal intrinsic scatter of $0.13\pm 0.02$.

Given the self-similarity of bulk gas mass and temperature measurements (Section~\ref{sec:results}), as well as gas density and temperature profiles \citep{Mantz1509.01322}, at radii of $\sim0.15$--1\,$r_{500}$, the strongly non-similar slope of the $Y$--$\Yx$ relation is challenging to explain. One consideration is the fact that the \Planck{} catalog information does not account for relativistic corrections to the SZ effect. Since we have temperature measurements for every cluster considered here, we can estimate the effect of this omission assuming that the \Planck{} $Y$ measurements are dominated by their 143\,GHz channel \citep{Chluba1205.5778, Chluba1211.3206}. Applying these corrections on a cluster-by-cluster basis and re-fitting the scaling relation results in an increase of the slope of only $\sim0.03$. The correction could be more significant if higher-frequency data are influential in the \Planck{} measurements, but we cannot be more quantitative with the information in hand.

Given that \Planck{} does not resolve any of the clusters in the sample considered here, another possible explanation is that the form of the pressure profile used to derive Compton $Y$ values from the \Planck{} data is not universal, but rather varies systematically with mass. At radii smaller than 0.15\,$r_{500}$, which are excluded from the X-ray temperature measurement, we might expect changes in the frequency of cool cores with mass to have an impact; however, we do not observe any systematic trends in surface brightness peakiness or dynamical relaxation with mass or $Y$ within this sample. Trends with mass of the parameters used to describe the cluster pressure profile, e.g.\ the slope at large radii, are seen in simulations \citep{Battaglia1109.3709, Le-Brun1501.05666} and data \citep{Ramos-Ceja1412.6023, Sayers1605.03541}, although not at a sufficient level to explain the measured $Y$--$M$ slope. 

Unlike \Planck{}, ground-based telescopes that measure the SZ effect typically resolve distant clusters; however, sensitivity at radii $\gtsim r_{500}$ suffers due to field of view limitations and atmospheric filtering. They thus probe complementary spatial scales, although the lack of complete coverage means that assumptions about the form of the pressure profile are still necessary to extract a $Y$ value from the data. SPT \citep{Andersson1006.3068} and APEX-SZ \citep{Bender1404.7103} measurements provide $Y$--$\Yx$ slopes that are steeper than ours, but consistent with both our results and unity within statistical uncertainties (respectively, $0.90\pm0.17$ and $0.88^{+0.16}_{-0.11}$). The $Y$--$\Yx$ slope of $0.84\pm0.07$ measured from Bolocam data \citep{Czakon1406.2800} is in good agreement with ours, although we note that the Bolocam slope applies to measurements within $r_{2500}$ rather than $r_{500}$. While we cannot draw any firm conclusions from these comparisons, an improved understanding of cluster pressure profile shapes and their dependence on mass and redshift is clearly important for cosmology with SZ cluster surveys, going forward.

\section{Conclusion} \label{sec:conclusion}

We present \Chandra{} and ROSAT X-ray measurements of gas mass, temperature and luminosity, and weak lensing measurements of total mass, for a subset of the X-ray flux limited cluster sample used to obtain cosmological constraints in Paper~\pfour{}, as well as constraints on the scaling relations of these quantities. Our scaling relation model is fitted using the methods described in Paper~\pfour{}, which account for selection effects, dependences on the underlying cosmology, and correlated measurement uncertainties. These are the first results from such a complete analysis which directly incorporate weak lensing data to provide the absolute cluster mass calibration.

Our constraints on the X-ray scaling relations are in good agreement with previous work in the literature. We find the power-law slopes of the soft-band X-ray luminosity, center-excised temperature, and gas mass with total mass to be, respectively, $\beta_{1_\ell}=1.34\pm0.05$, $\beta_{1,t}=0.62\pm0.04$ and $\beta_{1,\mgas}=1.007\pm0.012$. The latter two are consistent with the predictions of spherical collapse in the absence of additional heating or cooling (the self-similar model), which respectively predicts values of $2/3$ and unity. The soft-band luminosity--mass slope, however, significantly departs from the self-similar prediction of $\sim0.9$ for hot clusters. These results reinforce the picture of the ICM as simple and self-similar throughout most of the volumes of massive clusters, with non-gravitational heating and cooling processes driving departures from self similarity only in cluster cores.

The data reveal a positive correlation between the intrinsic scatters of luminosity and temperature at fixed mass, $\rho_{\ell t}=0.56\pm0.10$. This trend can be understood in terms of the dynamical states of the clusters, with temperature tracing mass most reliably in clusters that are close to virial equilibrium, and being relatively lower in merging clusters where energy in bulk motions has not yet virialized. Residuals in luminosity from the nominal scaling relations are straightforwardly dominated by the presence or absence of dense, bright cores, which occur preferentially, though not exclusively, in dynamically relaxed clusters. Comparison of the temperature and luminosity residuals to X-ray morphological and radio indicators of cluster dynamics supports this picture.

With follow-up X-ray data for $>100$ clusters at redshifts $0<z<0.5$ now in the analysis, we begin to see evidence for evolutionary departures from self similarity in the scaling relations. Interestingly, however, the strongest such evidence is not for additional evolution in the mean scaling relations but in the size of their intrinsic scatters. We find tentative ($\sim2\sigma$) evidence for negative evolution (decreasing with $z$) of the scatter in luminosity and positive evolution (increasing with $z$) of the scatter in temperature at fixed mass. Again, potential astrophysical explanations for these trends are readily available, namely the ongoing development of bright, dense cores in a fraction of clusters in the case of luminosity, and evolution in the rate of major mergers among massive clusters in the case of temperature. The possibility of constraining these aspects of cluster astrophysics strongly motivates the expansion of this analysis to higher redshifts and samples selected through different (non-X-ray) methods.

Although Compton $Y$ and optical richness are not yet incorporated into our complete cosmology+scaling relation analysis, we present empirical scaling relations of these observables with our X-ray and lensing measurements. We measure a normalization and slope of SDSS redMaPPer richness--mass relation that agree well with recent results based on stacked weak lensing, and an intrinsic scatter that is in line with previous estimates for X-ray clusters. Using Compton $Y$ measurements from \Planck{}, we find that the slope of the $Y$--mass and $Y$--$\Yx$ relations depart significantly from self-similarity; this is consistent with previous indications of a mass-dependent offset between weak lensing masses and masses estimated from \Planck{} $Y$ when assuming an approximately self-similar scaling. This trend is particularly surprising in light of the self-similar results we obtain for the gas mass and temperature scaling relations from X-ray data. Since the slope and evolution of the $Y$--$M$ relation directly influence inferences about $\Omegam$ and dark energy from SZ cluster surveys (through the shape of the mass function), understanding the source of this discrepancy should be a priority.

\section*{Acknowledgements}
We thank Jack Sayers and Sunil Golwala for useful discussions. We acknowledge support from the U.S. Department of Energy under contract number DE-AC02-76SF00515, and from the National Aeronautics and Space Administration under Grant No.\ NNX15AE12G issued through the ROSES 2014 Astrophysics Data Analysis Program.

\def \araa {ARA\&A}
\def \aj {AJ}
\def \aar {A\&AR}
\def \apj {ApJ}
\def \apjl {ApJL}
\def \apjs {ApJS}
\def \asl {Adv. Sci. Lett.} 
\def \mnras {MNRAS}
\def \nat {Nat}
\def \pasj {PASJ}
\def \pasp {PASP}
\def \science {Sci}
\def \gca {Geochim.\ Cosmochim.\ Acta}
\def \npa {Nucl.\ Phys.\ A}
\def \plb {Phys.\ Lett.\ B}
\def \prc {Phys.\ Rev.\ C}
\def \prd {Phys.\ Rev.\ D}
\def \prl {Phys.\ Rev.\ Lett.}
\def \jcap {J. Cosmology Astropart. Phys.} 
\def \physrep {Phys. Rep.} 
\def \aap {A\&A} 
\def \ijmpd {Int.\ J.\ Mod.\ Phys.\ D} 
\def \sjos {Scand.\ J.\ Statis.} 
\def \jrssb {J.\ Roy.\ Statist.\ Soc.\ B} 
\def \aapr {A\&AR} 
\def \ssr {Space Sci.\ Rev.}


\begin{table*}
  \begin{center}
    \setcounter{table}{0}\caption{continued}
    \vspace{-3ex}
\begin{tabular}{lccrccc}
      \hline
      Name & RA & Dec & Exp. & WtG & Rel. & OBSIDs \\
      \hline
      Abell~1914           &  14:26:02.894  &  +37:49:28.14    &  23.2   &  ~        &  ~        &  542,3593                                                                                                                                                                                                                                                                                                                                                                                              \\
Abell~2009           &  15:00:19.591  &  +21:22:10.27    &  18.1   &  ~        &  ~        &  10438                                                                                                                                                                                                                                                                                                                                                                                                 \\
Abell~2029           &  15:10:56.001  &  +05:44:40.31    &  118.9  &  ~        &  $\surd$  &  891,4977,6101,10434,10435,10436,10437                                                                                                                                                                                                                                                                                                                                                                 \\
\hline
   \end{tabular}
  \end{center}
\end{table*} 

\begin{table*}
  \begin{center}
    \setcounter{table}{0}\caption{continued}
    \begin{tabular}{lccrccc}
      \hline
      Name & RA & Dec & Exp. & WtG & Rel. & OBSIDs \\
      \hline
      MACS~J0011.7$-$1523  &  00:11:42.844  &  $-$15:23:20.42  &  49.2   &  ~        &  $\surd$  &  3261,6105                                                                                                                                                                                                                                                                                                                                                                                             \\
MACS~J0035.4$-$2015  &  00:35:26.494  &  $-$20:15:47.65  &  18.8   &  ~        &  ~        &  3262                                                                                                                                                                                                                                                                                                                                                                                                  \\
MACS~J0152.5$-$2852  &  01:52:34.464  &  $-$28:53:37.32  &  14.7   &  ~        &  ~        &  3264                                                                                                                                                                                                                                                                                                                                                                                                  \\
\hline
   \end{tabular}
  \end{center}
\end{table*}

\begin{table*}
  \begin{center}
    \setcounter{table}{1}\caption{continued}
    \begin{tabular}{lcr@{$\pm$}lr@{$\pm$}lr@{$\pm$}lr@{$\pm$}lc}
      \hline
      Name & $z$ & \multicolumn{2}{c}{$r_{500}$} & \multicolumn{2}{c}{$\Mgas$} & \multicolumn{2}{c}{$kT$} & \multicolumn{2}{c}{$L$} & $\Mlens$ \\
      &  & \multicolumn{2}{c}{(Mpc)} & \multicolumn{2}{c}{$(10^{14}\Msun)$} & \multicolumn{2}{c}{(keV)} & \multicolumn{2}{c}{$(10^{44}\erg\second^{-1})$} & $(10^{15}\Msun)$ \\
      \hline
      Abell~2597           &  0.085  &     0.97  &         0.04  &     0.36  &         0.04  &   4.44   &       0.12  &   4.21   &       0.05  &    ---                     \\
Abell~478            &  0.088  &     1.44  &         0.06  &     1.15  &         0.14  &   7.56   &       0.21  &   14.55  &       0.34  &    ---                     \\
Abell~2142           &  0.089  &     1.54  &         0.05  &     1.41  &         0.14  &   12.19  &       0.42  &   10.62  &       0.28  &    ---                     \\
Abell~3695           &  0.089  &     1.23  &         0.06  &     0.72  &         0.11  &   6.63   &       0.50  &   2.54   &       0.07  &    ---                     \\
Abell~3921           &  0.094  &     1.17  &         0.04  &     0.62  &         0.06  &   6.57   &       0.27  &   3.04   &       0.05  &    ---                     \\
Abell~2244           &  0.097  &     1.38  &         0.06  &     1.03  &         0.14  &   6.29   &       0.17  &   8.60   &       0.56  &    ---                     \\
\hline
    \end{tabular}
  \end{center}
\end{table*}

\begin{table*}
  \begin{center}
    \setcounter{table}{1}\caption{continued}
    \begin{tabular}{lcr@{$\pm$}lr@{$\pm$}lr@{$\pm$}lr@{$\pm$}lc}
      \hline
      Name & $z$ & \multicolumn{2}{c}{$r_{500}$} & \multicolumn{2}{c}{$\Mgas$} & \multicolumn{2}{c}{$kT$} & \multicolumn{2}{c}{$L$} & $\Mlens$ \\
      &  & \multicolumn{2}{c}{(Mpc)} & \multicolumn{2}{c}{$(10^{14}\Msun)$} & \multicolumn{2}{c}{(keV)} & \multicolumn{2}{c}{$(10^{44}\erg\second^{-1})$} & $(10^{15}\Msun)$ \\
      \hline
      RX~J0638.7$-$5358    &  0.227  &     1.36  &         0.05  &     1.13  &         0.12  &   11.14  &       1.03  &   11.14  &       0.20  &    ---                     \\
Abell~1763           &  0.228  &     1.61  &         0.06  &     1.88  &         0.21  &   9.09   &       0.67  &   9.42   &       0.21  &    ---                     \\
Abell~2219           &  0.228  &     1.79  &         0.06  &     2.55  &         0.26  &   12.80  &       0.36  &   16.89  &       0.19  &    $1.45^{+0.20}_{-0.19}$  \\
RX~J0118.1$-$2658    &  0.228  &     1.27  &         0.05  &     0.92  &         0.10  &   8.46   &       0.77  &   5.74   &       0.15  &    ---                     \\
RX~J0220.9$-$3829    &  0.228  &     1.00  &         0.04  &     0.45  &         0.05  &   6.01   &       0.61  &   3.84   &       0.13  &    ---                     \\
Abell~2111           &  0.229  &     1.29  &         0.05  &     0.95  &         0.11  &   9.07   &       0.70  &   5.05   &       0.12  &    ---                     \\
\hline
    \end{tabular}
  \end{center}
\end{table*}

\label{lastpage}
\end{document}